\documentclass[a4paper,fleqn]{cas-dc}
\ExplSyntaxOn
\cs_set:Npn \__first_footerline: {}
\ExplSyntaxOff
\usepackage[T1]{fontenc}
\usepackage{times}  

\usepackage{threeparttable} 
\usepackage{siunitx}
\usepackage{subcaption}
\usepackage{adjustbox}
\usepackage{wrapfig}
\usepackage[
    style=authoryear,
    maxcitenames=1,
    maxbibnames=99,
    backend=biber,
    sorting=nyt
]{biblatex}

\DeclareCiteCommand{\citeblue}
  {\usebibmacro{prenote}}
  {\href{https://doi.org/\thefield{doi}}{\printtext[bibhyperref]{\printnames{labelname}~\printfield{year}}}}
  {\multicitedelim}
  {\usebibmacro{postnote}}

\DeclareCiteCommand{\customcite}
  {\usebibmacro{prenote}}
  {\href{https://doi.org/\thefield{doi}}{\textcolor{blue}{(\printnames{labelname}, \printfield{year})}}}
  {\multicitedelim}
  {\usebibmacro{postnote}}

\newcommand{\blueeqref}[1]{\hyperref[#1]{\textcolor{blue}{\ref*{#1}}}}
\usepackage{xspace}
\usepackage{etoolbox}
\newcommand{\tsc}[1]{\csdef{#1}{\textsc{\lowercase{#1}}\xspace}}
\tsc{WGM} \tsc{QE} \tsc{EP} \tsc{PMS} \tsc{BEC} \tsc{DE}
\usepackage{caption}
\DeclareCaptionFont{rm}{\rmfamily}
\let\printorcid\relax
\shortauthors{}
\begin{document}

\let\WriteBookmarks\relax
\def\floatpagepagefraction{1}
\def\textpagefraction{.001}
\shorttitle{\normalsize \textrm{Geotechnical Property Characterization during Microtunneling: A Bayesian
Sequential Inference Approach}}

\title [mode=title]{Geotechnical Property Characterization during Microtunneling: A Bayesian Sequential Inference Approach}


\author[1]{Honglai Peng}
\cormark[1]

\author[2]{Rodrigo Borela}

\author[3]{Kimberlie Staheli}

\author[4]{J. David Frost}

\affiliation[1]{organization={School of Computational Science and Engineering \& School of Civil and Environmental Engineering, Georgia Institute of Technology}, 
                city={Atlanta}, 
                state={GA}, 
                country={United States}}

\affiliation[2]{organization={School of Computing Instruction, Georgia Institute of Technology}, 
                city={Atlanta}, 
                state={GA}, 
                country={United States}}

\affiliation[3]{organization={Staheli Trenchless Consultants, Inc.}, 
                state={WA}, 
                country={United States}}

\affiliation[4]{organization={School of Civil and Environmental Engineering, Georgia Institute of Technology}, 
                city={Atlanta}, 
                state={GA}, 
                country={United States}}


\begin{abstract}
Microtunneling is a widely used trenchless construction technique that minimizes surface disruption, reduces environmental impact, and maintains precise alignment. Real-time operational measurements, such as jacking force, offer valuable insights into subsurface characterization. However, reliable geotechnical-property characterization remains challenging due to measurement uncertainty, spatial variability, and the coupled relationship between operational measurements and ground conditions. Measured jacking forces entangle face resistance, pipe–soil friction, and measurement error, allowing different parameter combinations to produce similar observations and making the inference ill-posed. This study develops a Bayesian sequential inverse framework to characterize subsurface conditions by inferring geotechnical properties during microtunneling. The method integrates site-investigation information and monitored data to regularize parameter inference and sequentially update parameter distributions as tunneling progresses, while explicitly quantifying uncertainty and capturing spatial variability. The proposed method is first evaluated using synthetic experiments to assess parameter recovery under varying noise levels and sensitivity to modeling errors. Then, benchmark tests across three unlubricated drive segments demonstrate more stable parameter estimates than genetic algorithm and particle swarm optimization methods, consistent with field expectations, while achieving comparable force reconstruction. Finally, the framework is evaluated using a lubricated microtunneling case, showing that the inference responds to observations beyond the initial assumptions, while revealing compensation among parameters and force components that limits force decomposition and parameter identification under lubricated conditions. The framework provides a probabilistic basis for interpreting microtunneling measurements for ongoing subsurface characterization, with potential benefits for construction decision-making and risk management.
\end{abstract}

\begin{highlights}
\item Bayesian sequential inversion infers soil properties from monitored jacking forces.
\item Site-investigation priors regularize the ill-posed inversion of jacking-force data.
\item Three field drives show that force agreement alone cannot identify soil parameters.
\item Benchmarks against GA and PSO show more stable, physically plausible inferences.
\item A lubricated case shows adaptability, parameter compensation, and inference limits.

\end{highlights}


\maketitle

\section{Introduction}
Microtunneling, a trenchless construction technology, is increasingly adopted in geotechnical engineering for installing pipelines beneath existing infrastructure with minimal surface disruption. Compared with traditional open-cut methods, it is particularly advantageous in densely populated urban environments, where minimizing surface settlement, traffic disruption, and environmental impact is critical. Uncertainties in subsurface conditions can directly affect construction safety and performance, making continuous subsurface characterization along the tunnel alignment essential for reliable microtunneling operations.

Microtunnel Boring Machines (MTBMs) provide continuous support at the tunnel face to balance earth and groundwater pressures, helping to limit surface settlement and maintain construction safety and precision (\citeblue{french2010microtunneling, phillips2019instrumentation, cheng2023face}). Modern machines are able to continuously record operational measurements such as jacking force, torque, face pressure, and advancement rate through onboard sensors, which offer valuable insights about the interaction between the machine and the surrounding ground. However, effectively leveraging such data for geotechnical inference remains challenging due to high measurement uncertainty, limited data volumes relative to spatial variability, and noise introduced by operational and environmental disturbances. Geotechnical site-characterization data exhibit characteristics described by MUSIC: multivariate, uncertain and unique, sparse, and incomplete (\citeblue{doi:10.1061/AJRUA6.0000992, phoon2019managing, yuen2021bayesian}). Moreover, geotechnical-property inference from jacking-force measurements is inherently ill-posed. In unlubricated drives, these measurements entangle face resistance at the cutting head, frictional resistance along the pipe–soil interface, and measurement errors. The force components depend on the empirical face-resistance factor, residual interface friction coefficient, residual friction angle, and total soil unit weight. Therefore, without a suitable mathematical framework and physical constraints, different combinations of these parameters can produce similar force responses, leading to non-unique and potentially physically implausible inferences.

Bayesian methods provide a framework for tackling ill-posed inverse problems by incorporating prior knowledge and explicitly quantifying uncertainty (\citeblue{mariethoz2010bayesian, calvetti2018inverse}). In geotechnical engineering, these methods support inference under uncertainties associated with geometry, measurement errors, and approximations in deterministic models. A Bayesian machine learning method is introduced for constructing site-specific multivariate probability distribution models by combining site-specific and generic data according to their availability (\citeblue{ching2019constructing}). Machine learning (ML) has also seen growing adoption in microtunneling applications (\citeblue{sheil2020machine}), with probabilistic machine learning models applied to jacking-force prediction in microtunneling projects (\citeblue{sheil2020prediction, sheil2022prediction, yu2024bayesian, yu2024prediction}). 

Despite these advances, existing studies have primarily focused on forward prediction tasks, such as jacking-force estimation, while subsurface characterization through the inverse inference of spatially varying soil properties along the tunnel drive remains largely unexplored as a primary research objective. To address this gap, this study develops a Bayesian sequential inverse framework to characterize subsurface conditions by inferring geotechnical properties during microtunneling. The framework integrates site-investigation information and monitored jacking-force data to regularize parameter inference and sequentially update parameter distributions with explicit uncertainty quantification. 
The contributions of this study are as follows:
\begin{itemize}
    \item We develop a Bayesian sequential inverse framework to characterize subsurface conditions by inferring geotechnical properties during microtunneling, integrating site-investigation information with monitored jacking forces to regularize the ill-posed inference and explicitly quantify parameter uncertainty throughout the tunneling process.
    
    \item We introduce a data preprocessing pipeline that integrates Savitzky–Golay filtering, Seasonal-Trend decomposition using Loess (STL), and quantile–quantile (Q–Q) analysis to improve data quality.
    \item We evaluate the framework using controlled synthetic datasets to assess parameter recovery, robustness to measurement noise, and sensitivity to modeling errors.
    \item We benchmark the framework against Genetic Algorithm (GA) and Particle Swarm Optimization (PSO) using real-world data from unlubricated microtunneling segments, demonstrating that site-investigation-based initial distributions regularize the ill-posed inference and yield more stable, physically plausible, and interpretable geotechnical-property inferences while achieving comparable force-reconstruction performance.
    \item We extend the evaluation to a lubricated microtunneling segment to assess the framework’s adaptability to altered pipe–soil interaction and examine its limitations in force decomposition and parameter identification under lubricated conditions.
\end{itemize}

\section{Bayesian inference approach}\label{bayesian inference approach}

\begin{figure}[htbp]
    \centering
    \includegraphics[width=0.95\linewidth]{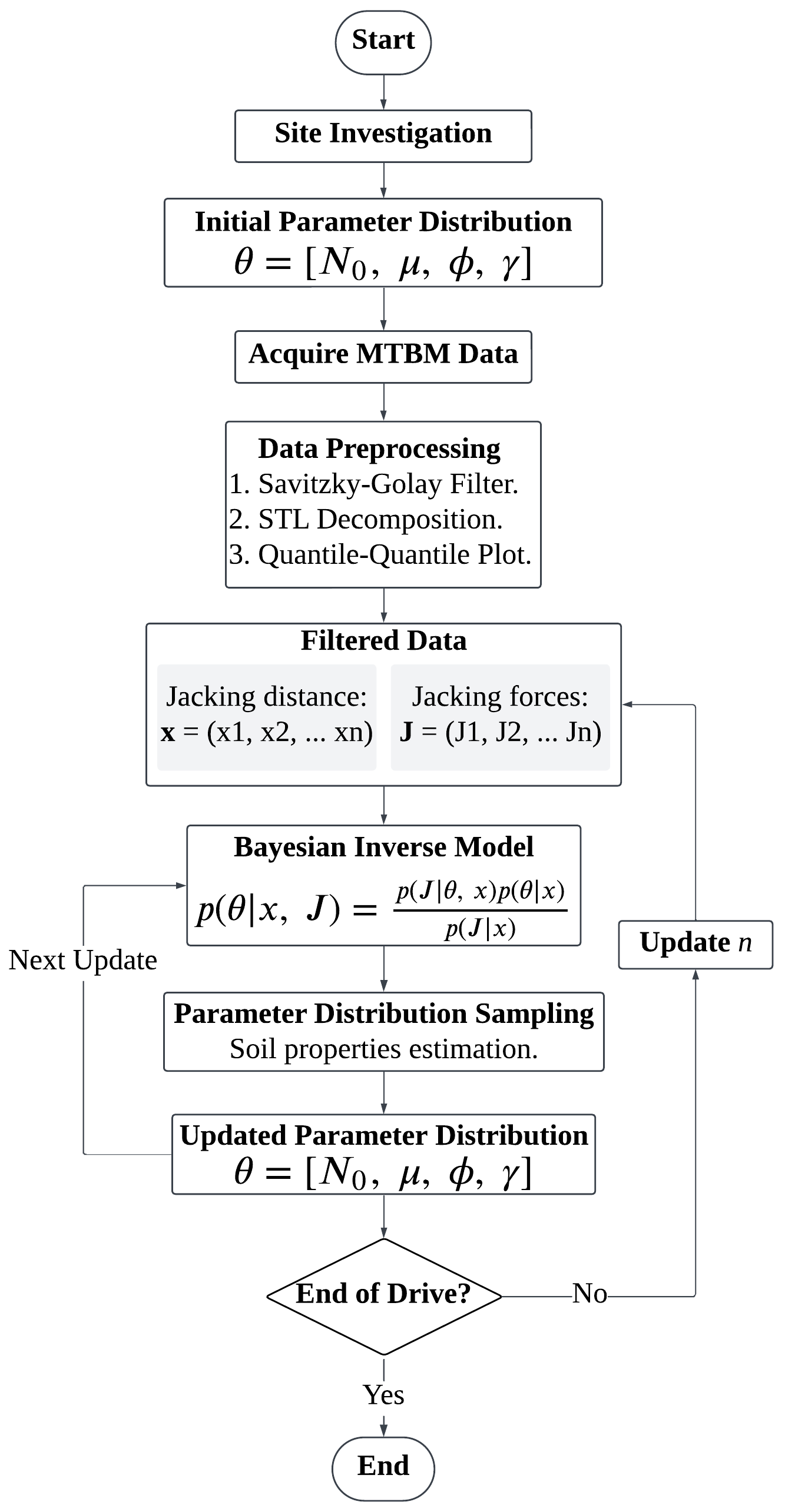} 
    \caption{\normalsize \textrm{The workflow for Bayesian sequential inference of geotechnical properties during microtunneling.}}
    \label{flowchart}
\end{figure}

In this study, we develop a Bayesian inference framework consisting of four main components, as shown in Fig.~\blueeqref{flowchart}: preprocessing of MTBM monitoring data, construction of initial parameter distributions from site-investigation information, formulation of the likelihood function, and sampling of updated parameter distributions. 

 
\subsection{Data Preprocessing} 
We propose a preprocessing pipeline to improve the quality of data used for Bayesian inference. While the Savitzky–Golay filter is effective in suppressing high-frequency noise in tunneling data and has been widely adopted in previous studies (\citeblue{sheil2022prediction, yu2024bayesian, yu2024prediction}), it is not well suited for handling outliers. We therefore incorporate Seasonal-Trend decomposition using Loess (STL) and quantile--quantile (Q--Q) plots, which are commonly used in time-series analysis to assess residual behavior and identify potential outliers (\citeblue{cleveland6seasonal, hyndman2018forecasting, gnanadesikan1968probability})

The preprocessing begins with the Savitzky–Golay filter to reduce high-frequency noise. STL decomposition is then applied to isolate the residual component by removing any underlying trend or seasonal patterns. Since no significant trend or periodicity was identified in the tunneling datasets, the STL parameters were set to minimal values to retain abrupt variations in the data. This simplification avoids extensive parameter tuning and allows subsequent adjustments to focus on the initial filtering step when necessary.

Potential outliers are assessed using a Q–Q plot of the STL residuals. Under well-behaved noise conditions, the residuals are expected to approximately follow a normal distribution. If substantial deviations from normality are observed, the Savitzky–Golay filter parameters are revisited and adjusted iteratively until the residuals more closely approximate normality. Once the residuals fall within an acceptable range, the Q–Q plot is further used to identify and remove statistical outliers. This procedure improves data quality and is particularly suitable for small tunneling datasets, in which irregular and non-ideal observations can adversely affect subsequent inference.

\subsection{Initial Parameter Distribution Constructions} 
In tunneling projects, contractors typically conduct site investigations, including borehole investigations, to characterize subsurface conditions and support the selection of suitable cutting equipment. In this study, the investigation results are used to construct the initial parameter distributions $P(\theta)$ for the Bayesian model. Following common practice in geotechnical engineering (\citeblue{li2016assessment, zheng2018embankment}), the parameters $\theta$ are modeled as lognormal random variables by assuming that their logarithms, denoted by $\boldsymbol{\theta}_{\log}=\log(\boldsymbol{\theta})$, follow a multivariate normal (MVN) distribution. This formulation ensures strictly positive parameter values while allowing to capture the uncertainties and correlations among the parameters:
\begin{equation}
p(\boldsymbol{\theta}_{\log}) =
\operatorname{MVN}(\boldsymbol{\eta},\boldsymbol{\Lambda})
\label{E1}
\end{equation}
where $\eta$ and $\Lambda$ are the mean vector and covariance matrix in log space, respectively. The parameter vector $\theta$ is defined as:
\begin{equation}
\boldsymbol{\theta}=[N_0,\ \mu,\ \phi,\ \gamma]
\label{E2}
\end{equation}
where $N_0$ is the empirical face resistance factor, $\mu$ is the residual interface friction coefficient, $\phi$ is the residual friction angle,  and $\gamma$ is the total unit weight.

\subsection{Likelihood Function}\label{deterministic}
In our study, we adopt deterministic forward models to construct the likelihood function. In microtunneling, the jacking force at a given location is modeled as:
\begin{equation}
\boldsymbol{J}
=
\boldsymbol{F}(\boldsymbol{\theta},\boldsymbol{x})
+
\boldsymbol{\epsilon}
\label{E3}
\end{equation}

where $\boldsymbol{J}=(J_1,J_2,\dots,J_n)$ denotes the vector of observed jacking forces corresponding to the jacking distances $\boldsymbol{x}=(x_1,x_2,\dots,x_n)$. The vector $\boldsymbol{F}(\boldsymbol{\theta},\boldsymbol{x})$ contains the forces predicted by the deterministic forward model for the parameter vector \(\boldsymbol{\theta}\), as formulated through Eqs. \blueeqref{E4}–\blueeqref{E8}. The residual term $\boldsymbol{\epsilon}$ represents measurement errors.

At a given jacking distance, the modeled total jacking force is decomposed into face resistance and frictional resistance along the pipe–soil interface:
\begin{equation}
F_T = F_{\mathrm{face}} + F_{\mathrm{friction}}
\label{E4}
\end{equation}
where $F_T$ denotes the total jacking force, $F_{\mathrm{face}}$ is the total resistance acting at the cutting head, and the $F_{\mathrm{friction}}$ is the frictional resistance mobilized along the pipe--soil interface.

\noindent The total face resistance force exerted at the TBM cutterhead is defined as:
\begin{equation}
F_{\mathrm{face}} =
f_{\mathrm{face}}\,
\pi\left(\frac{D_{\mathrm{TBM}}}{2}\right)^2
\label{E5}
\end{equation}
The unit face resistance is estimated using the empirical formulation proposed by \citeblue{japan2000pipe} and \citeblue{shou2010frictional}:
\begin{equation}
f_{\mathrm{face}}
=
10.0 \times 1.32 \cdot \pi \cdot D_{\mathrm{TBM}} \cdot N_0
\label{E6}
\end{equation}
where $f_{\mathrm{face}}$ is the unit face resistance $(\mathrm{kN}/\mathrm{m}^2)$, $D_{\mathrm{TBM}}$ is the TBM cutterhead diameter $(\mathrm{m})$, and $N_0$ is an empirical face-resistance factor. Typical values of $N_0$ are 1.0 for clayey soils, 2.5 for sandy soils, and 3.0 for gravelly soils.

The total frictional resistance along the pipe string is defined as:
\begin{equation}
F_{\mathrm{friction}} =
f_{\mathrm{friction}}\,\pi D_p L_p
\label{E7}
\end{equation}
The unit skin friction is estimated using the empirical expression proposed by \citeblue{staheli2006jacking}:
\begin{equation}
f_{\mathrm{friction}}
=
\mu
\frac{
\gamma \left(D_p/2\right)
\cos\left(45^\circ+\phi/2\right)
}{
\tan\phi
}
\label{E8}
\end{equation}
where $f_{\mathrm{friction}}$ is the unit skin friction resistance $(\mathrm{kN}/\mathrm{m}^2)$, $\mu$ is the residual pipe–soil interface friction coefficient, $\gamma$ is the total soil unit weight $(\mathrm{kN}/\mathrm{m}^3)$, $\phi$ is the residual friction angle of the soil $(^\circ)$, $D_p$ is the pipe diameter $(\mathrm{m})$, and $L_p$ is the pipe length $(\mathrm{m})$.

The measurement errors $\epsilon$ in Eq. \blueeqref{E3} are assumed to be statistically independent across observation locations and are modeled as normally distributed. These errors may result from electromagnetic interference, mechanical vibrations, temperature-dependent sensor response, and signal-transmission losses during data acquisition (\citeblue{beck1998updating, jaynes2003probability, box2011bayesian}).

Given the deterministic forward model $\boldsymbol{F}(\boldsymbol{\theta},\mathbf{x})$ and the assumption of normally distributed measurement errors, the likelihood function can be formulated within the Bayesian inference framework. Bayes’ theorem provides the mathematical basis for updating the initial parameter distribution $p(\boldsymbol{\theta})$ using the observed jacking forces $\mathbf{J}$ at the corresponding distances $\mathbf{x}$:
\begin{equation}
p(\boldsymbol{\theta}\mid\mathbf{J},\mathbf{x})
=
\frac{
p(\mathbf{J}\mid\boldsymbol{\theta},\mathbf{x})p(\boldsymbol{\theta})
}{
p(\mathbf{J}\mid\mathbf{x})
}
\label{E9}
\end{equation}
Assuming that the measurement errors are independent across observation locations, the joint likelihood is expressed as the product of the individual likelihoods:
\begin{equation}
p(\mathbf{J}\mid\boldsymbol{\theta},\mathbf{x})
=
\prod_{i=1}^{n}
p(J_i\mid\boldsymbol{\theta},x_i)
\label{E10}
\end{equation}

\subsection{Parameter Distribution Sampling}
The term $p(\mathbf{J}\mid\mathbf{x})$ in Eq. \blueeqref{E9} represents the marginal likelihood, also referred to as the model evidence. It serves as the normalization constant that ensures the updated parameter distribution integrates to one and is defined as:
\begin{equation}
p(\mathbf{J}\mid\mathbf{x})
=
\int
p(\mathbf{J}\mid\boldsymbol{\theta},\mathbf{x})
p(\boldsymbol{\theta})
\,d\boldsymbol{\theta}.
\label{E11}
\end{equation}
In Bayesian inference, when the prior and likelihood form a conjugate pair, the posterior distribution can be derived analytically in closed form (\citeblue{gelman1995bayesian, murphy2012machine}). In this study, however, the deterministic jacking-force model introduces nonlinear relationships among the parameters, resulting in a likelihood that is not conjugate to the parameter distribution. Consequently, the updated distribution does not admit a closed-form solution. In addition, evaluating the marginal likelihood in Eq. \blueeqref{E11} requires integration over the parameter space, which is generally intractable for this nonlinear geotechnical inverse problem. To address these challenges, we employ Markov chain Monte Carlo (MCMC) methods to generate samples from the updated distribution without directly evaluating the marginal likelihood (\citeblue{van2018simple}). MCMC constructs a Markov chain whose stationary distribution corresponds to the target distribution, enabling the generation of representative samples through iterative updates (\citeblue{gamerman2006markov, gilks1995markov}).
The high dimensionality of the parameter space and the strong correlations among parameters motivate the use of the No-U-Turn Sampler (NUTS), a gradient-based extension of Hamiltonian Monte Carlo (HMC) that improves sampling efficiency and convergence in complex posterior landscapes (\citeblue{hoffman2014no}).

\section{Benchmark algorithms}
Two population-based optimization algorithms, Particle Swarm Optimization (PSO) and Genetic Algorithm (GA), are used as benchmark methods for comparison with the proposed MCMC-based Bayesian sequential inference approach.
\subsection{Particle Swarm Optimization}
Particle Swarm Optimization (PSO) is a population-based optimization method that searches the solution space by iteratively updating a set of candidate solutions according to their objective-function values (\citeblue{kennedy1995particle}). PSO has been extensively applied in tunneling-related research (\citeblue{hajihassani2018applications}). Representative applications include enhancing machine-learning models for jacking-force prediction (\citeblue{zhou2023prediction}), optimizing XGBoost hyperparameters for forecasting tunnel deformation from rock–machine interaction data (\citeblue{bo2024prediction}), and combining PSO with Support Vector Machines (SVMs) to model nonlinear relationships between tunneling parameters and surrounding-ground displacements (\citeblue{jiang2011feedback}).

\begin{figure*}[h]
    \centering
    \begin{subfigure}[t]{0.72\linewidth}
        \centering
        \includegraphics[width=\linewidth]{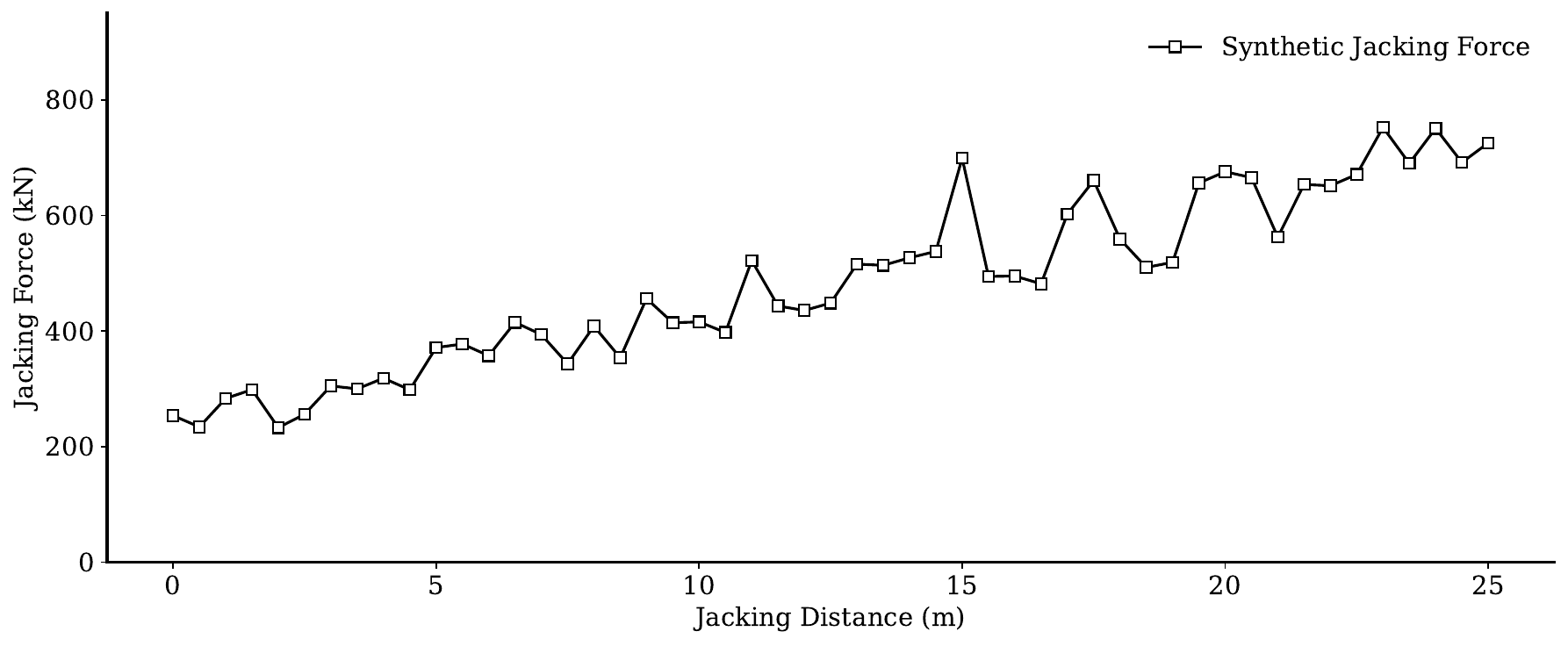}
        \phantomsubcaption
        \label{Synthetic Data}
        {\small \textrm{(a)}}
    \end{subfigure}
    \begin{subfigure}[t]{0.72\linewidth}
        \centering
        \includegraphics[width=\linewidth]{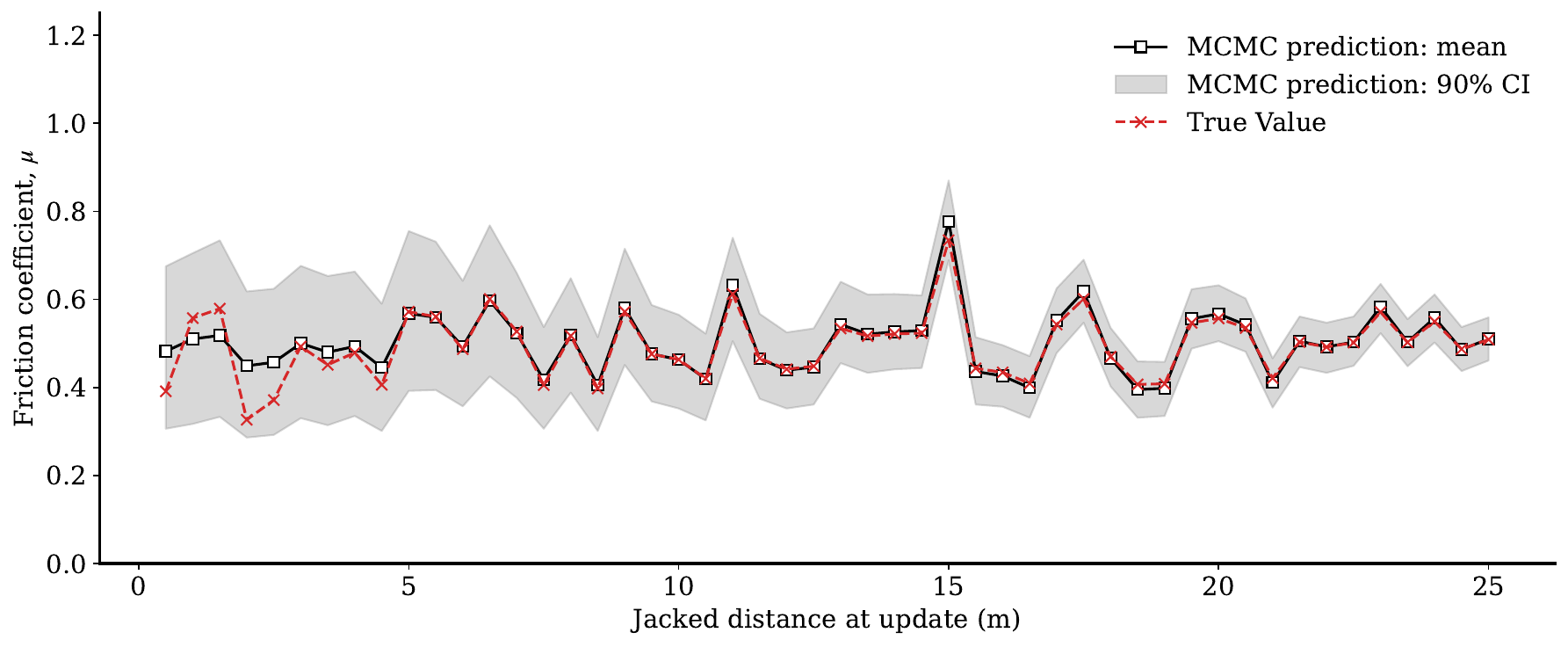}
        \phantomsubcaption
        \label{mu_100_20}
        {\small \textrm{(b)}}
    \end{subfigure}


    \begin{subfigure}[t]{0.72\linewidth}
        \centering
        \includegraphics[width=\linewidth]{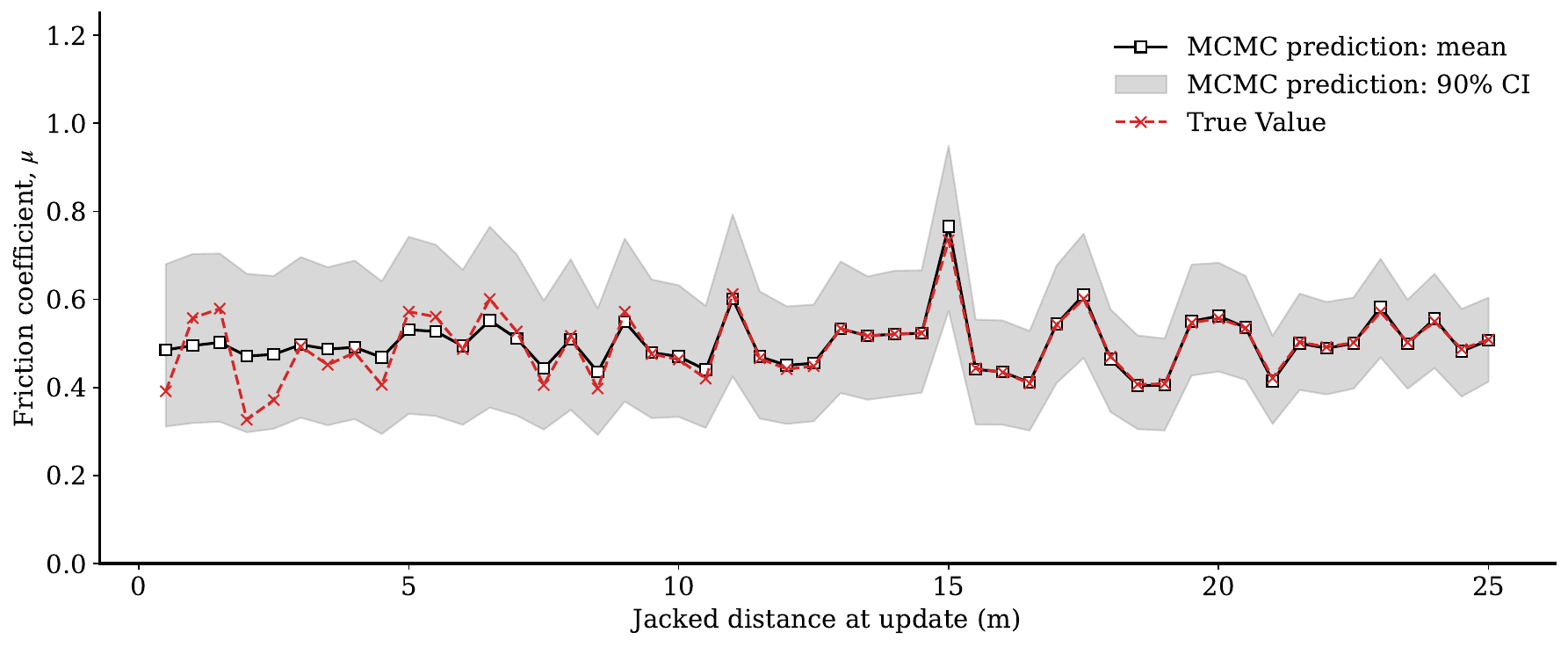}
        \phantomsubcaption
        \label{mu_100_50}
        {\small \textrm{(c)}}
    \end{subfigure}
    
    \begin{subfigure}[t]{0.72\linewidth}
        \centering
        \includegraphics[width=\linewidth]{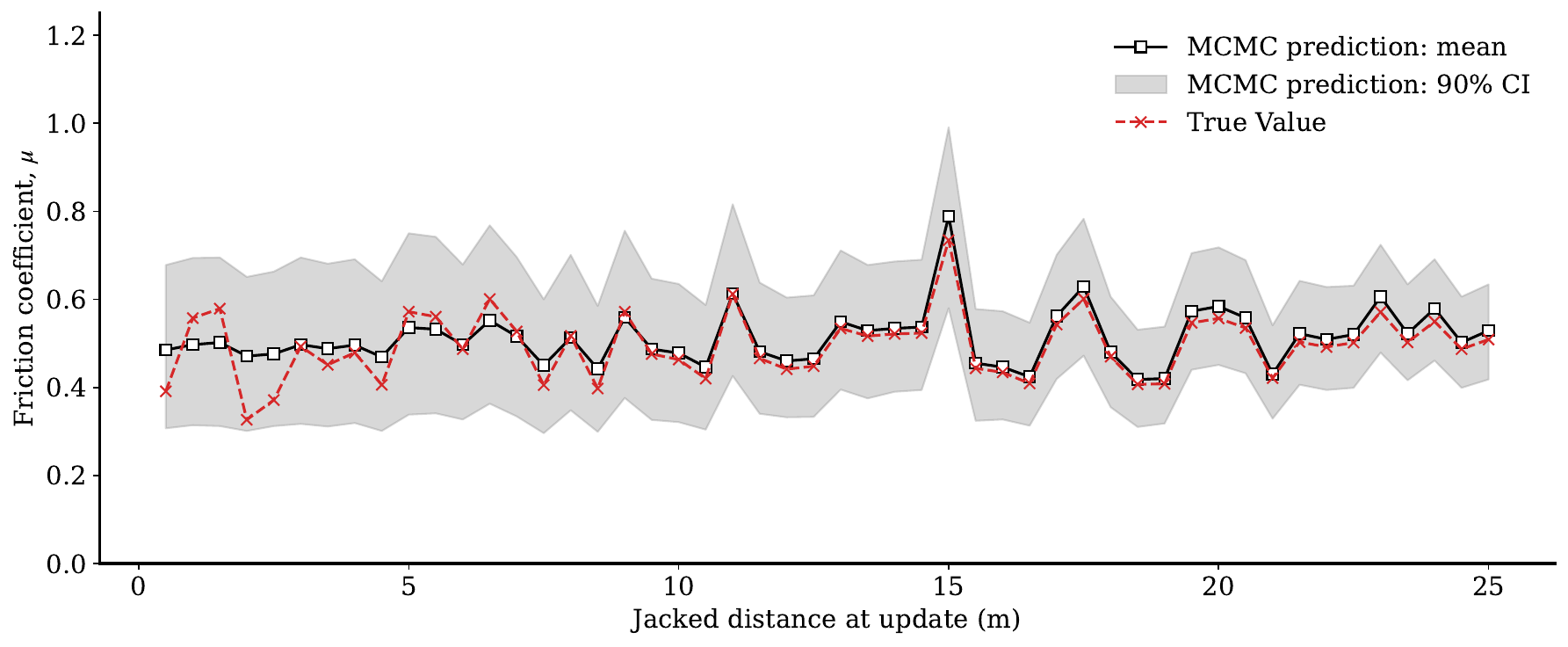}
        \phantomsubcaption
        \label{model_error_100_50}
        {\small \textrm{(d)}}
    \end{subfigure}

    \caption{\normalsize \textrm{Experimental evaluation using synthetic data: (a) example of the synthesized total jacking force profile along the tunnel drive; (b) sequential inference of the friction coefficient $\mu$ under measurement noise $\epsilon \sim \mathcal{N}(100, 20)$; (c) sequential inference of the friction coefficient $\mu$ under measurement noise $\epsilon \sim \mathcal{N}(100, 50)$; and (d) sequential inference of the friction coefficient $\mu$ with additional structural perturbation in the forward model.}}
    \label{Synthetic Experimental Results}
\end{figure*}

\subsection{Genetic Algorithm}
Genetic Algorithm (GA), inspired by the principles of natural evolution, is a very classic heuristic search method (\citeblue{holland1992genetic}) and has been widely applied to engineering design optimization problems (\citeblue{andrab2017review}). In geotechnical applications, GA has been used to identify the constitutive parameters of the Mohr–Coulomb model from the field measurements, demonstrating its applicability to complex parameter-identification problems (\citeblue{levasseur2008soil}). Hybrid genetic algorithm–neural network (GA–NN) approaches have also been used to construct prediction intervals for tunnel settlement by optimizing neural-network weights with a GA-based objective function that accounts for model uncertainty and noise variance (\citeblue{feng2022enhanced}). In addition, GA-based methods have been applied to back-analyze tunnel responses and improve parameter estimation by aligning model predictions with monitoring data (\citeblue{vardakos2012parameter}).

\section{Evaluation on synthetic data}
To systematically evaluate the proposed Bayesian sequential inference framework, we first apply it to controlled synthetic data before considering the real-world case studies. The synthetic experiments aim to assess the ability of the framework to recover known parameters, quantify the robustness to controlled measurement noise, and examine its sensitivity to controlled model perturbations.
\subsection{Synthetic Data Generation}
The synthetic datasets are generated using soil-property values derived from the site-investigation records for the Alvarado Sewer project as the mean parameter values. The corresponding coefficients of variation (COVs) are obtained from the literature and summarized in Table \blueeqref{prior parameters}. The logarithms of the parameters $\boldsymbol{\theta}=[N_0,\ \mu,\ \phi,\ \gamma]$ are jointly sampled from a multivariate normal distribution. The simulated tunnel drive spans 25 m, corresponding to the unlubricated segment of the Alvarado Sewer case. At each jacking distance, the total jacking force, face resistance, and pipe–soil friction are calculated using the deterministic forward model described in Eqs. \blueeqref{E4}–\blueeqref{E8}.
\begin{table*}[!htbp]
    \centering
    {\fontfamily{ptm}\selectfont  
    \caption{\normalsize \textrm{Initial parameter distributions for the microtunneling case studies}}
    \label{prior parameters}
    \renewcommand{\arraystretch}{1.2}
    \begin{tabular}{lccccc}
        \toprule
        \multirow{2}{*}{Parameter} & \multicolumn{2}{c}{Alvarado Sewer} & \multirow{2}{*}{Morris Avenue} & \multirow{2}{*}{Coefficient of Variation} & \multirow{2}{*}{Distribution} \\
        \cmidrule(lr){2-3}
        & JP3-RP4 & JP4-RP4 \\
        \midrule
        \textsuperscript{a} Residual Soil friction angle $\phi$ (\si{\degree}) & 31 & 30 & 32 & 0.075 & Lognormal \\
        \textsuperscript{b} Total Soil unit weight $\gamma$ (\si{kN/m^3}) & 21 & 20 & 17 & 0.05 & Lognormal \\
        \textsuperscript{c} Empirical face resistance factor $N_0$ & 2.5 & 2.5 & 2.5 & 0.2 & Lognormal \\
        \textsuperscript{d} Residual Interface friction coefficient $\mu$ & 0.49 & 0.45 & 0.59 & 0.86 & Lognormal \\
        \textsuperscript{e} Model and measurement error $\epsilon$ (kN) & 100 & 100 & 100 & 0.2 & Normal \\
        \bottomrule
    \end{tabular}

    \vspace{0.5em}
    \begin{minipage}{0.95\textwidth}
    \footnotesize
    \textsuperscript{a,b} Mean values of both cases are from site investigations and (\citeblue{staheli2006jacking}). COV value is from (\citeblue{harr1987reliability, kulhawy1992evaluation}). \\
    \textsuperscript{c} Mean values are from (\citeblue{shou2010frictional}). COV from (\citeblue{sheil2022prediction}). \\
    \textsuperscript{d} Mean values from site investigations and (\citeblue{staheli2006jacking}). COV from (\citeblue{shou2010frictional}). \\
    \textsuperscript{e} Mean values from (\citeblue{o2020interpretation}). COV from (\citeblue{sheil2022prediction}).
    \end{minipage}
    } 
\end{table*}
\begin{figure*}[hbp]
    \centering
    \begin{subfigure}[t]{0.452\linewidth}
        \centering
        \includegraphics[width=\linewidth]{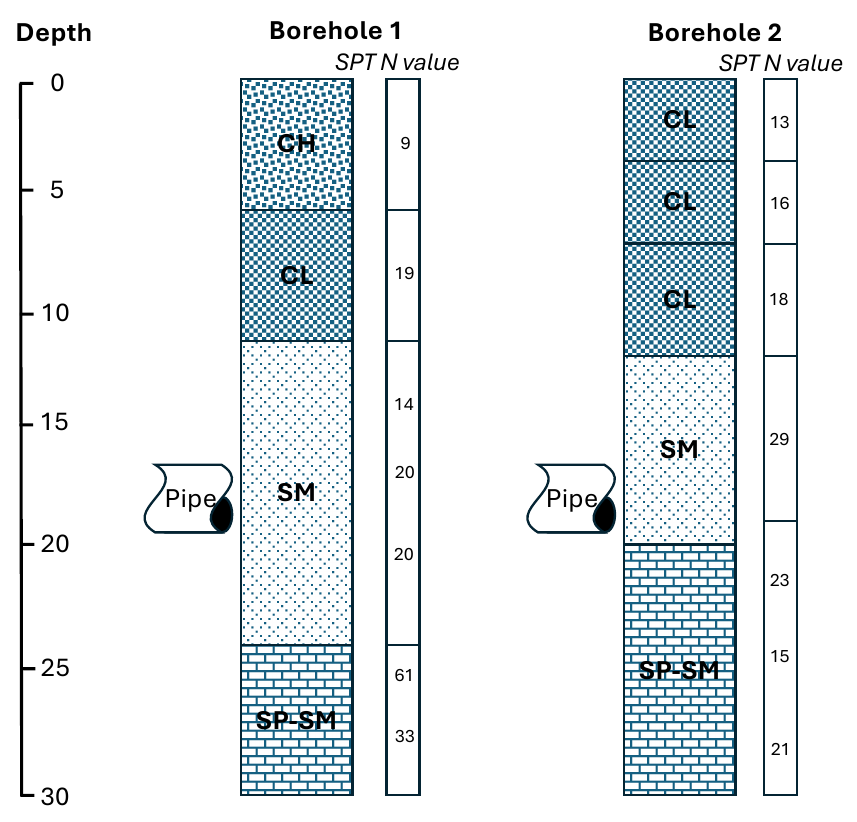}
        \label{JP3_4_BH}
        {\small \textrm{(a)}}
    \end{subfigure}
    \hfill
    \begin{subfigure}[t]{0.46\linewidth}
        \centering
        \includegraphics[width=\linewidth]{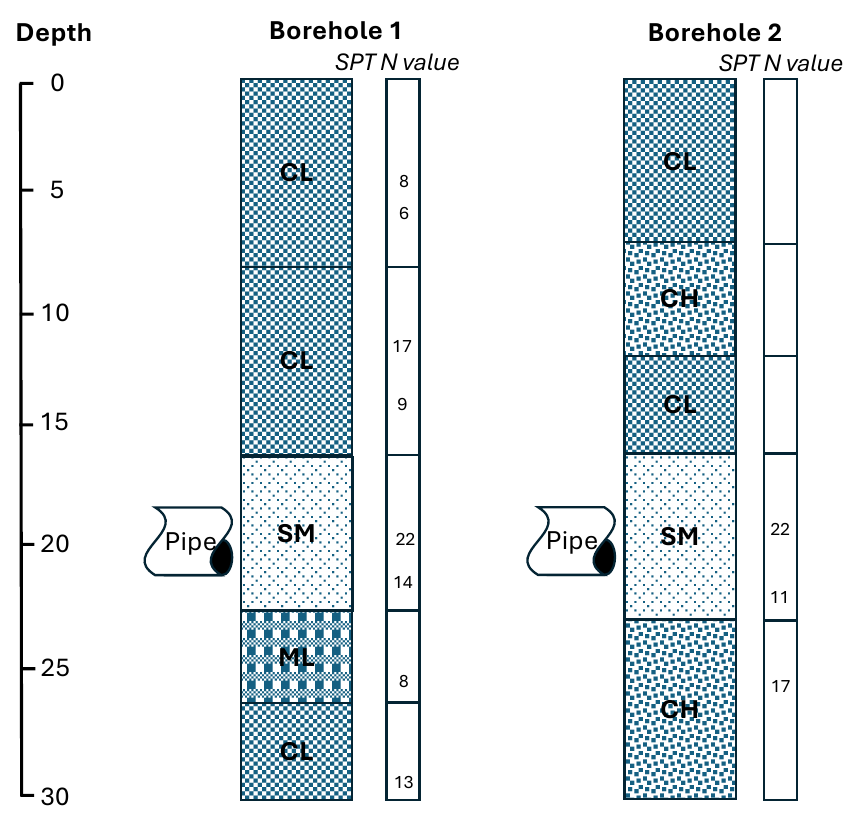}
        \label{JP4_4_BH}
        {\small \textrm{(b)}}
    \end{subfigure}
    \hfill
    \begin{subfigure}[t]{0.49\linewidth}
        \centering
        \includegraphics[width=\linewidth]{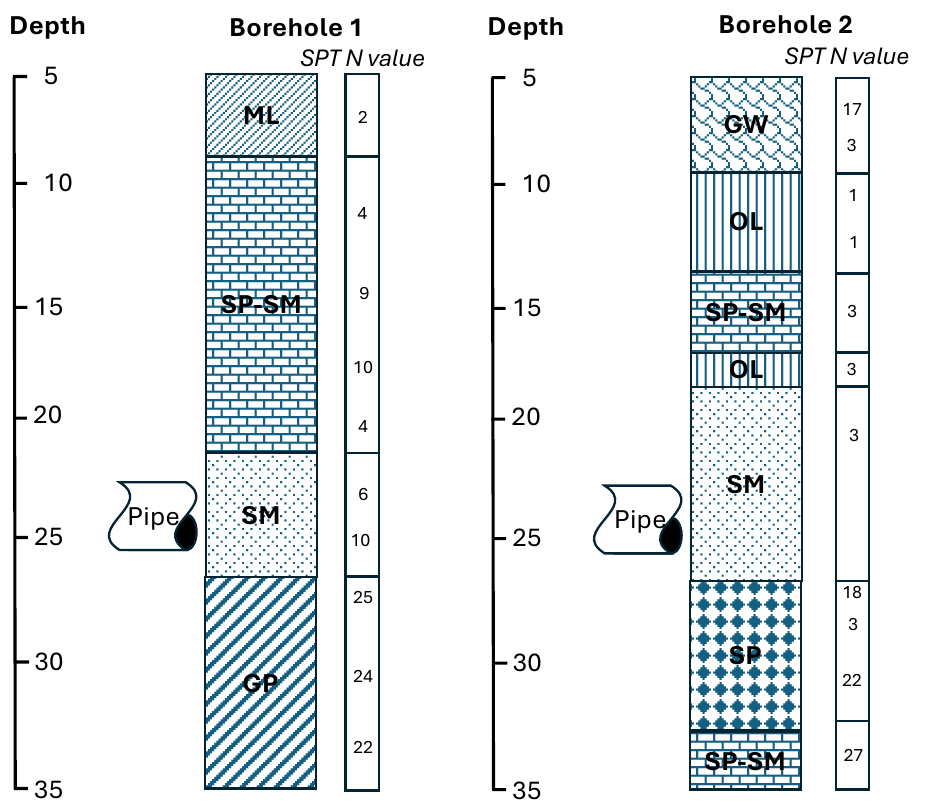}
        \label{Morris_BH}
        {\small \textrm{(c)}}
    \end{subfigure}
    \caption{\normalsize \textrm{Stratigraphic profiles: (a) the Alvarado Trunk Sewer JP3-RP4 project, (b) the Alvarado Trunk Sewer JP4-RP4 project, and (c) the Morris Avenue Tunnel project.}}
    \label{Borehole_Combined}
\end{figure*}
\subsection{Experimental Results}
Fig. \blueeqref{Synthetic Data} shows an example of the synthetic total jacking-force profile. The friction coefficient $\mu$ is selected as a representative parameter to examine sequential inference under controlled measurement noise and model perturbations, as shown in Figs. \blueeqref{mu_100_20}–\blueeqref{model_error_100_50}. 
Figs. \blueeqref{mu_100_20} and \blueeqref{mu_100_50} compare the inference results under noise settings of \(\epsilon \sim \mathcal{N}(100,20)\) and \(\epsilon \sim \mathcal{N}(100,50)\), respectively. In both cases, the inferred parameters initially deviate from the ground-truth value when few observations are available. As additional observations are incorporated, the inferences approach the true value and the associated uncertainty decreases. However, the higher-noise case in Fig. \blueeqref{mu_100_50} requires more observations for the inference to stabilize and converge. Increasing the noise level also raises the mean width of the 90\% credible intervals for \(\mu\) by 39.5\%, indicating greater uncertainty in the inferred friction coefficient. These results demonstrate that the Bayesian framework is able to reflect the measurement noise through the credible intervals while retaining parameter recovery.
Fig. \blueeqref{model_error_100_50} presents the inference results for a synthetic dataset with a deliberate model perturbation:
\begin{equation}\label{E12}
    f'_{\text{friction}} = f_{\text{friction}}(1+\Delta),
\end{equation}
\vspace{-0.8cm}
\begin{equation}\label{E13}
    F'_{\text{friction}} = f'_{\text{friction}}\pi D_pL_p,
\end{equation}
where $\Delta$ denotes the prescribed relative model perturbation. Compared with Fig. \blueeqref{mu_100_50}, the inferred estimates of $\mu$ show a systematic deviation from the ground-truth value, consistent with the imposed perturbation. 

The synthetic evaluation demonstrates parameter recovery, robustness to measurement noise, and sensitivity to model errors in the Bayesian inverse framework. The results also highlight that the framework is able to distinguish the effects of observational noise and model deficiencies under the controlled test conditions.

\section{Unlubricated case studies} \label{case}
In this section, we apply the proposed Bayesian sequential inference method to two microtunneling projects: the Alvarado Trunk Sewer and the Morris Avenue Tunnel. All selected tunnel segments are unlubricated, lie within relatively homogeneous soil layers, and were constructed below the groundwater table.

The Alvarado Trunk Sewer project comprises two separate drives: JP3–RP4 (Jacking Pit 3 to Reception Pit 4) and JP4–RP4 (Jacking Pit 4 to Reception Pit 4). For JP3–RP4, the method was applied to the initial 30 m from the jacking shaft, whereas for JP4–RP4, it was applied to the initial 25 m. For the Morris Avenue Tunnel, data from approximately 180 m to the end of the drive were used.
The stratigraphy files derived from the site investigations provided the initial information for subsurface characterization, as shown in Fig. \blueeqref{Borehole_Combined}. The subsurface conditions identified from these files informed the initial parameter distributions used in the Bayesian model, providing a physically meaningful starting point for subsequent probabilistic updating. The initial parameter-distribution values for each case are presented in Table \blueeqref{prior parameters}, including the assumed means, COVs, and probability distributions for each geotechnical parameter.

\begin{figure}[htbp]
    \centering
    \includegraphics[width=0.98\linewidth]{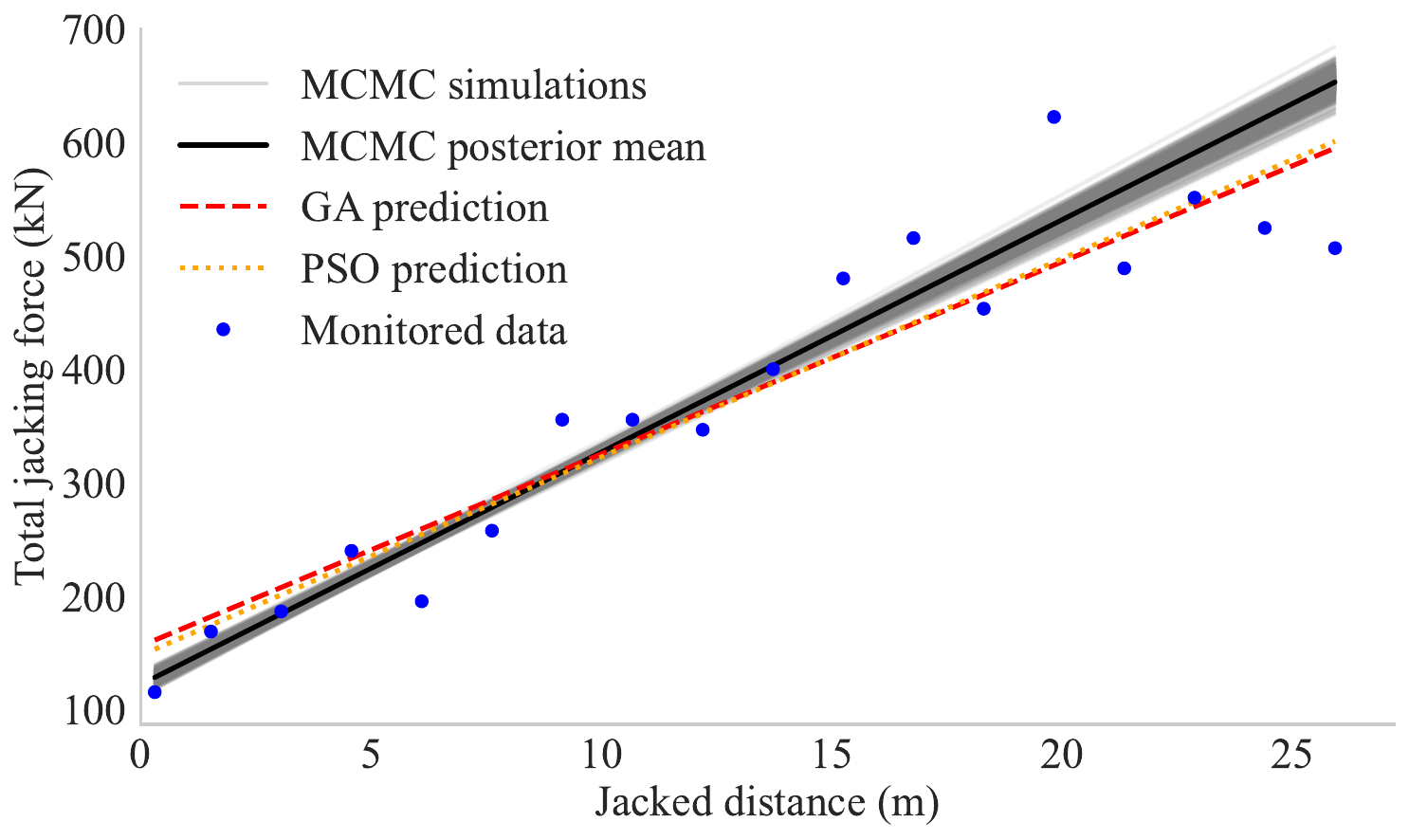} 
    \caption{\normalsize \textrm{Comparison of the reconstructed total jacking forces for the unlubricated segment of the Alvarado Trunk Sewer JP3–RP4 drive using MCMC-based Bayesian inference, GA, and PSO.}}
    \label{JP3_sim_comparison}
\end{figure}
\subsection{Alvarado Trunk Sewer JP3-RP4}\label{JP3_RP3}

The Alvarado Trunk Sewer JP3-RP4 project, located in Union City and Fremont, California, involved the installation of approximately 216.40 m of Polycrete pipe in a single continuous drive. The subsurface conditions for this case study were classified as medium-dense silty sand. The pipe had an outer diameter of 1.18 m and an inner diameter of 1.00 m. Tunneling was carried out using an Akkerman microtunneling machine equipped with a soft-ground cutting head. The cutterhead diameter was 1.22 m, providing an intentional diameter overcut of approximately 0.04 m to facilitate tunneling. The project incorporated a bentonite lubrication system supplied through a manifold connected to a port on the microtunneling machine. Lubrication was not activated at the beginning of the drive; bentonite injection began after approximately 30.50 m of advancement and continued till the end of the drive.

Fig. \blueeqref{JP3_sim_comparison} compares the jacking-force reconstructions using MCMC-based Bayesian inference, GA, and PSO. All methods reproduce the observed jacking-force trend with comparable agreement, yielding $R^2$ values between 0.85 and 0.88. The MCMC-based approach additionally quantifies uncertainty in the reconstructed forces through force simulations generated from posterior parameter samples. Despite the similarity in predicted forces, the three methods exhibit notable differences in how they update the underlying model parameters $\theta$, further illustrating that force agreement alone is insufficient for geotechnical-property identification.

\begin{figure*}[!htb]
    \centering
    \includegraphics[width=0.89\linewidth]{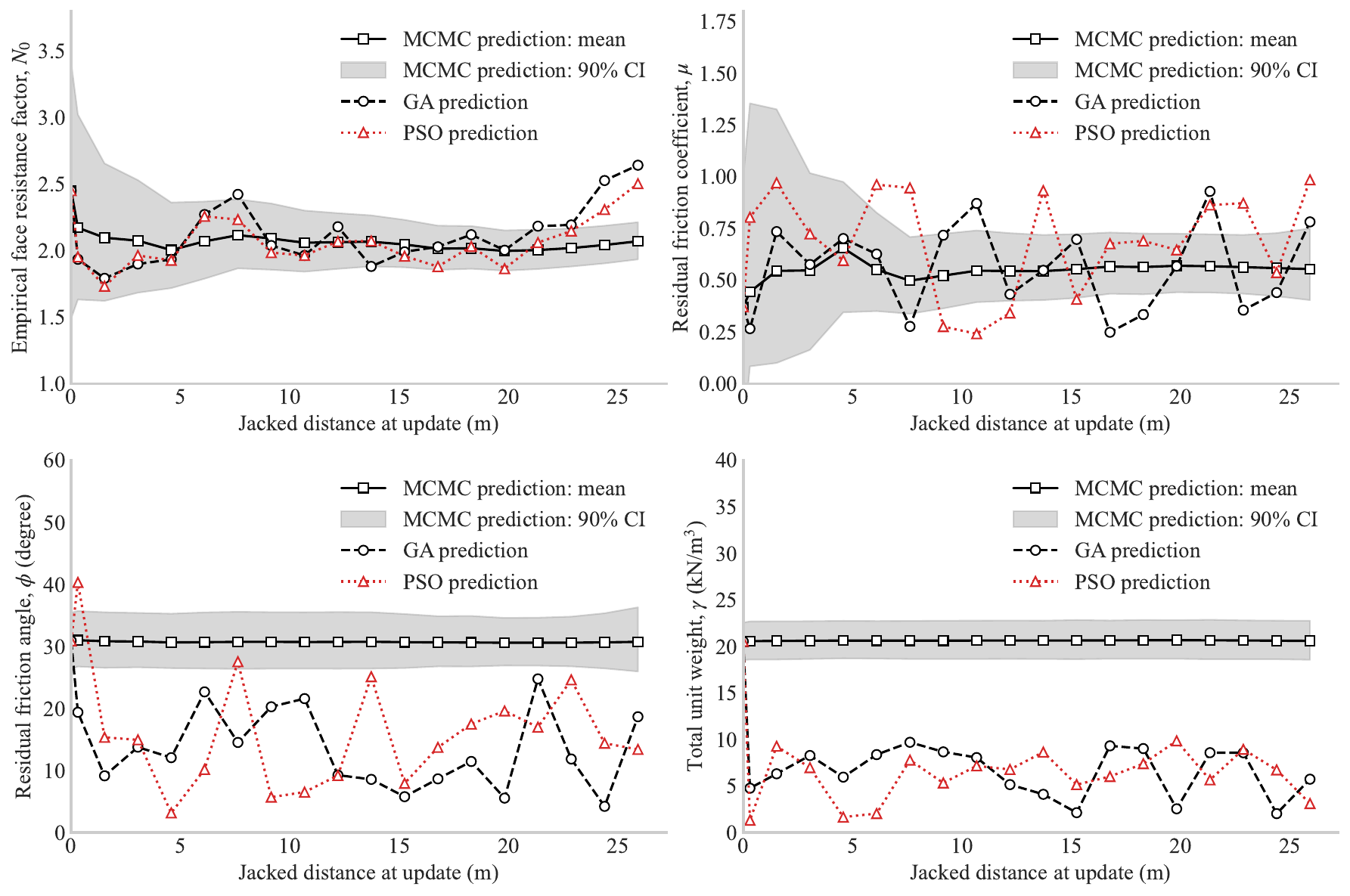} 
    \caption{\normalsize \textrm{Sequential estimates of the soil properties for the unlubricated segment of the Alvarado Trunk Sewer JP3–RP4 drive using MCMC-based Bayesian inference, GA, and PSO.}}
    \label{JP3_Comparison_Unlubricated}
\end{figure*}

The evolution of the parameter estimates obtained using the MCMC-based Bayesian framework, GA, and PSO is shown in Fig. \blueeqref{JP3_Comparison_Unlubricated}. For the MCMC-based results, the shaded gray bands represent the 90\% credible intervals (CI), indicating the uncertainty associated with the parameter inference along the drive.
In this study, the initial value of the empirical face-resistance factor was set to $N_0=2.5$, consistent with recommendations for sandy soils (\citeblue{japan2000pipe, shou2010frictional}), as discussed in Section \ref{deterministic}. All three methods show a slight decrease in \(N_0\) after the first update. The MCMC-based estimate subsequently stabilizes at approximately 2.2, with relatively narrow 90\% credible intervals. This value lies between the typical values reported for clean sandy soils ($N_0=2.5$) and clayey soils ($N_0=1.0$), which is consistent with the medium-dense silty-sand conditions identified for this case study. The result indicates that the sequential inference can adjust the initial estimate to reflect the observed jacking-force response while retaining a physically plausible value and quantifying the associated uncertainty along the drive. While the GA and PSO estimates follow broadly similar trends and converge toward similar central values, they exhibit greater variability across the sequential updates. This fluctuation suggests that although the GA and PSO can produce reasonable point estimates, their iterative search procedures are more sensitive to local data variations and lack the regularizing influence provided by the initial parameter distributions, which constrain the sequential updates while allowing the observations to update the inference.

\begin{figure*}[!htb]
    \centering
    \includegraphics[width=0.89\linewidth]{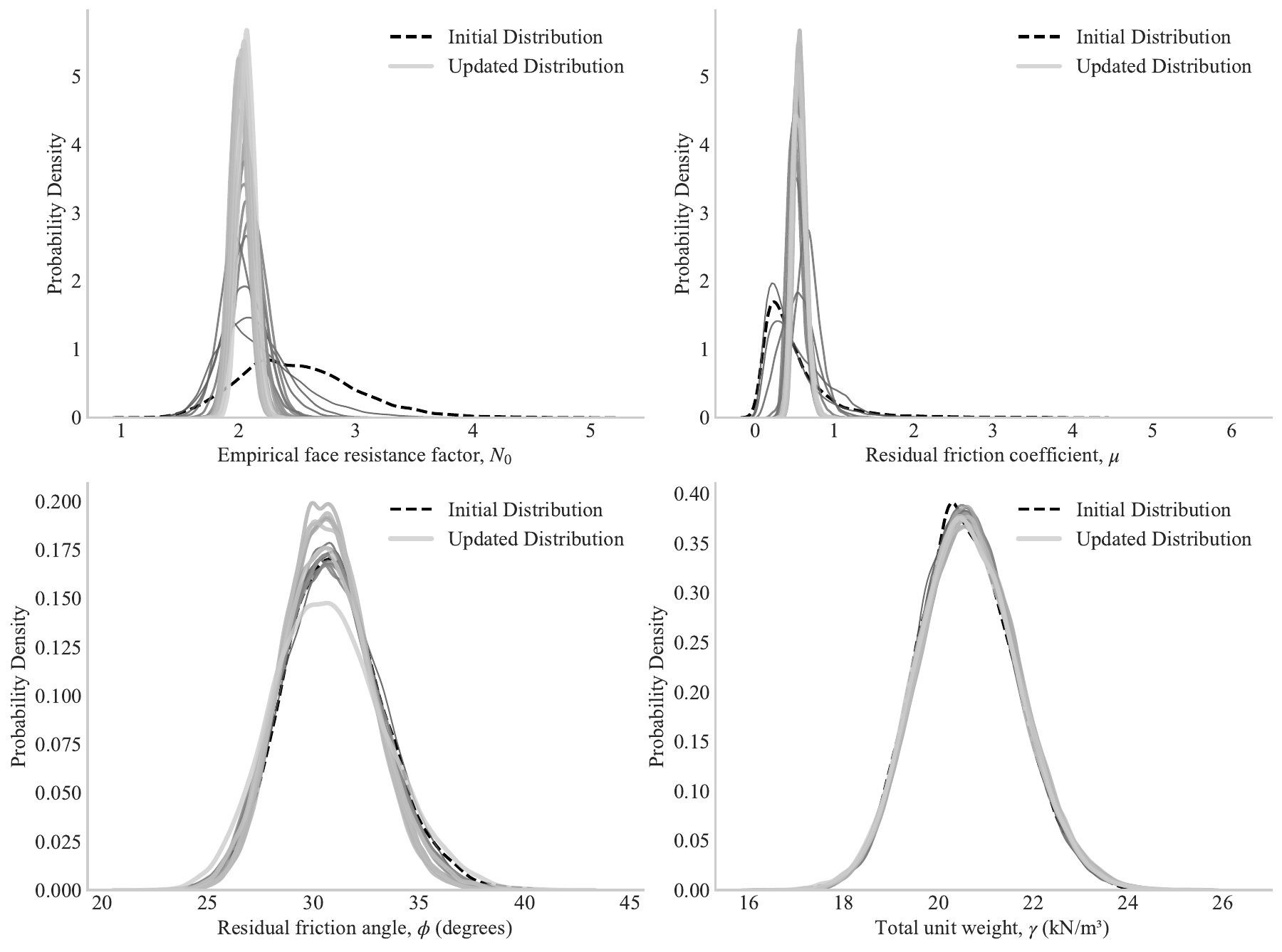} 
    \caption{\normalsize \textrm{Initial and sequentially updated parameter distributions for the unlubricated segment of the Alvarado Trunk Sewer JP3–RP4 drive.}}
    \label{JP3_prior_vs_posterior_updates}
\end{figure*}

For the residual interface friction coefficient $\mu$, the MCMC-based posterior estimates exhibit a gradual increase during the first four update steps, reaching a peak of approximately 0.6. Then, they gradually stabilize around 0.5, with narrowing 90\% credible intervals indicating reduced posterior uncertainty. Both GA and PSO show comparable upward trends during the initial updates. However, after approximately 5 meters into the drive, their trajectories begin to diverge, with PSO initially increasing while GA decreases and both methods fluctuating during subsequent updates. Although their average estimates are broadly comparable to the MCMC posterior mean, both methods display substantially greater variability in their estimates of $\mu$, suggesting greater sensitivity to changes in the observations. In addition, their point estimates do not provide the uncertainty quantification available through the Bayesian framework.

For the residual friction angle $\phi$ and total unit weight $\gamma$, the Bayesian approach produces updated estimates that remain close to the initial values, with 90\% credible intervals indicating the uncertainty associated with the estimates. In contrast, both GA and PSO exhibit a sharp decline in their estimates during the initial updates. Their estimates then remain consistently lower than the MCMC estimates and the site-investigation-based reference values, despite comparable force reconstructions, and fluctuate substantially during subsequent updates. These results highlight the sensitivity of GA and PSO to local data variations and show that force agreement alone does not establish physically meaningful parameter estimates. In contrast, the Bayesian framework carries the site-investigation-based initial distributions through the sequential updates to constrain these weakly identifiable parameters, while also quantifying the remaining uncertainty, resulting in more stable and physically meaningful estimates.

Fig. \blueeqref{JP3_prior_vs_posterior_updates} illustrates the evolution of the distributions of the four model parameters across sequential Bayesian updates, comparing them directly with their initial distributions. For $N_0$ and $\mu$, the updated distributions progressively become more concentrated around their updated means, reflecting the influence of additional observations and the sequential accumulation of information across updates. The progressive tightening of these distributions also indicates reduced uncertainty in the parameter estimates as more data are incorporated. In comparison, the updated distributions of $\phi$ and $\gamma$ remain closely aligned with their respective initial distributions, indicating limited adjustment during sequential inference, consistent with the relatively small COVs assigned to these parameters. Overall, the results illustrate how Bayesian sequential inference incorporates observational data to update parameter distributions, with the most evident reductions in uncertainty occurring for $N_0$ and $\mu$.

\subsection{Alvarado Trunk Sewer JP4-RP4}
The Alvarado Trunk Sewer JP4–RP4 project, located in Union City and Fremont, California, involved the installation of approximately 236.2 m of Polycrete pipe in a single microtunnel drive. This drive shared Reception Pit 4 with the adjacent JP3--RP4 drive discussed in Section \blueeqref{JP3_RP3}. At the pipeline elevation, the subsurface material was classified as medium-dense silty sand. The pipeline had an outer diameter of 1.18 m and an inner diameter of 1.00 m. Tunneling was carried out using an Akkerman microtunneling machine equipped with a soft-ground cutting head. The cutterhead diameter was approximately 1.22 m, providing an intentional diameter overcut of about 0.04 m to facilitate jacking. Bentonite lubrication was delivered through a 0.05-m pipeline connected to a port at the tail of the machine. Lubrication began approximately 25 m into the drive and continued till the end of the tunnel.
\begin{figure}[htbp]
    \centering
    \includegraphics[width=0.98\linewidth]{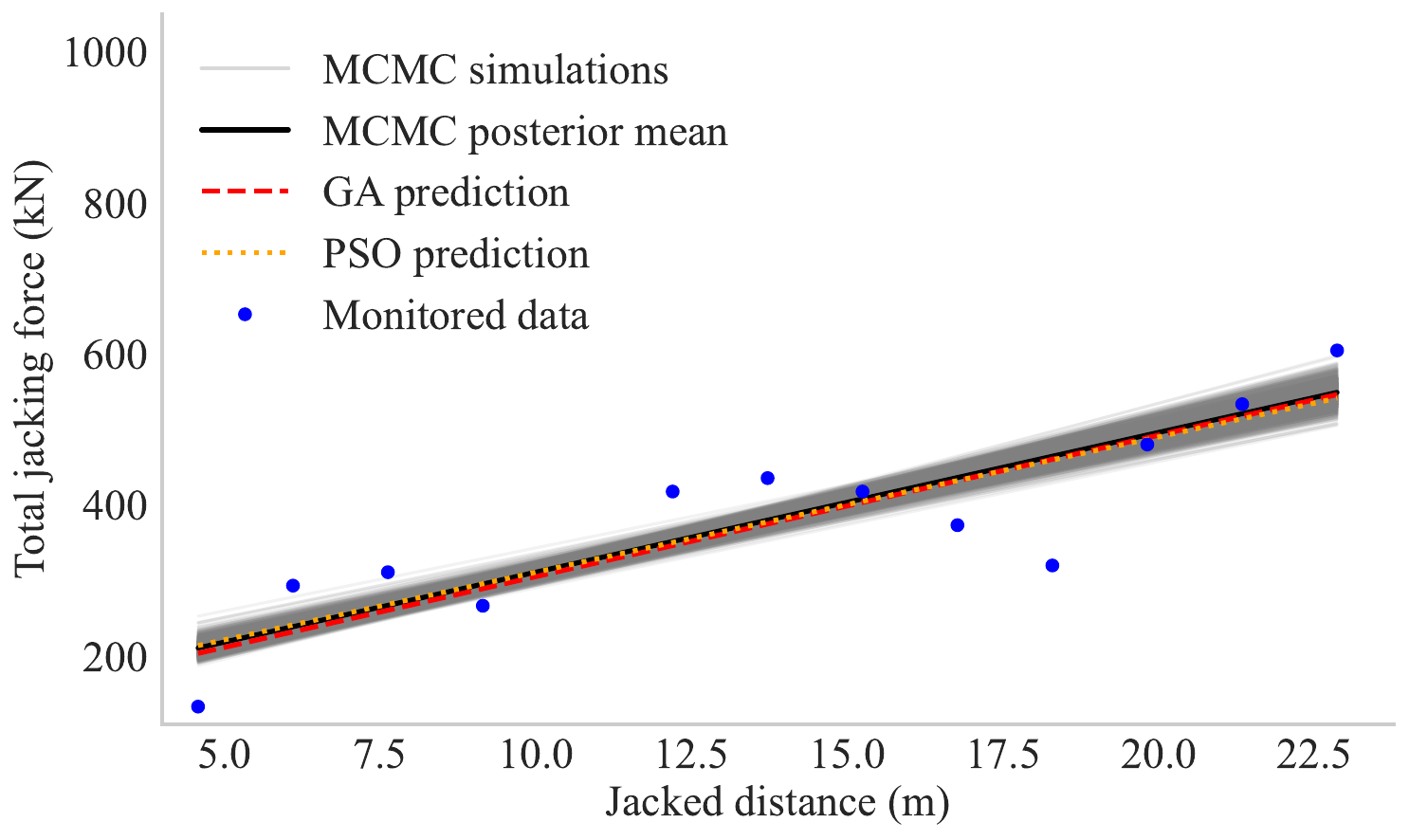} 
    \caption{\normalsize \textrm{Comparison of the reconstructed total jacking forces for the unlubricated segment of the Alvarado Trunk Sewer JP4–RP4 drive using MCMC-based Bayesian inference, GA, and PSO.}}
    \label{JP4_sim_comparison}
\end{figure}

\begin{figure*}[!htb]
    \centering
    \includegraphics[width=0.89\linewidth]{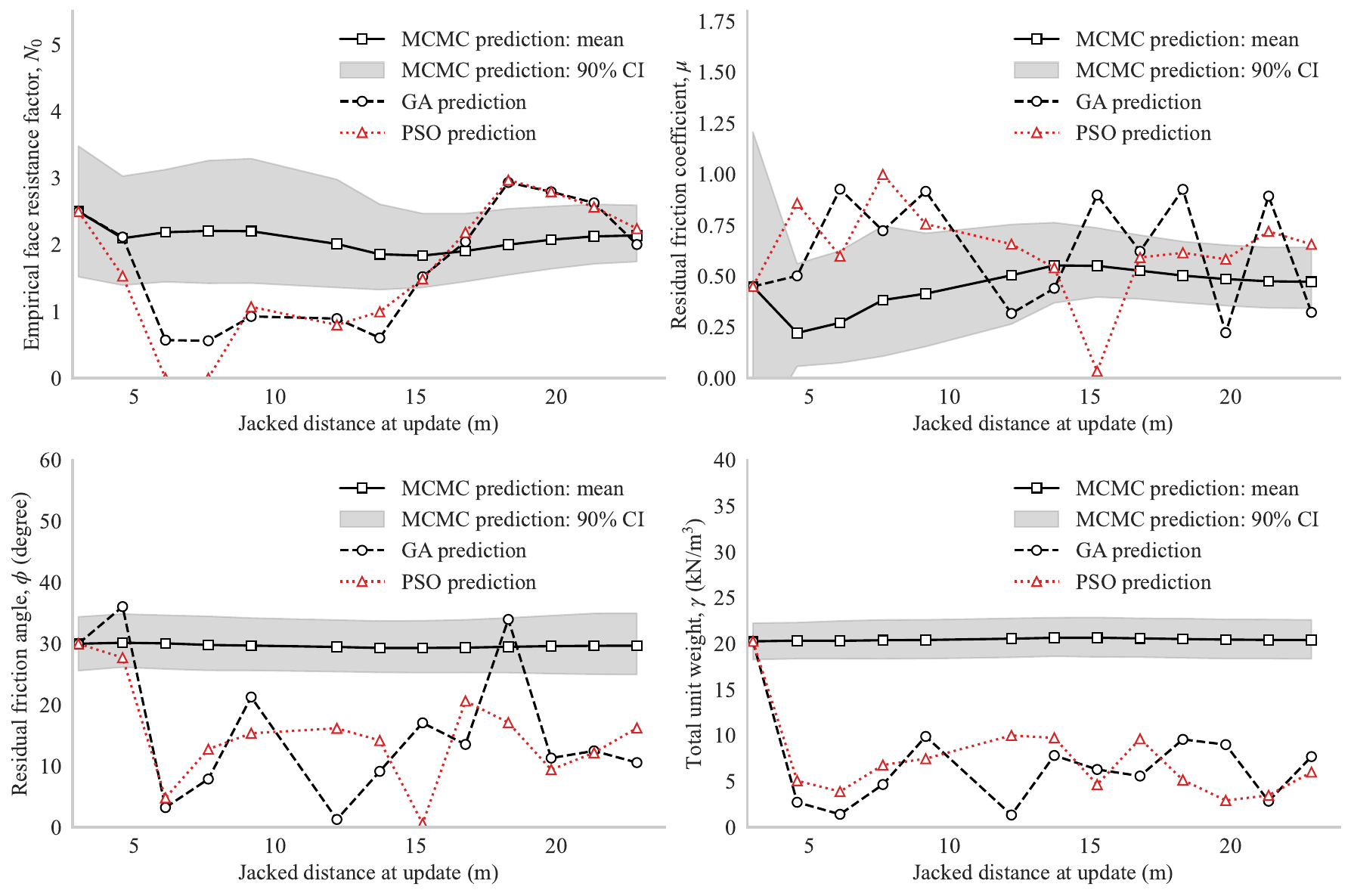} 
    \caption{\normalsize \textrm{Sequential estimates of the soil properties for the unlubricated segment of the Alvarado Trunk Sewer JP4–RP4 drive using MCMC-based Bayesian inference, GA, and PSO.}}
    \label{JP4_Comparison_Unlubricated}
\end{figure*}

Fig. \blueeqref{JP4_sim_comparison} presents the jacking-force reconstructions for the JP4--RP4 drive using all three methods. As with the JP3--RP4 results shown in Fig. \blueeqref{JP3_sim_comparison}, the methods produce comparable force reconstructions, with $R^2$ values ranging from 0.73 to 0.74 against the monitored data. However, their underlying parameter updates differ substantially, as shown in Fig. \blueeqref{JP4_Comparison_Unlubricated}, again illustrating that comparable force reconstruction can accompany different parameter estimates.

In this study, an initial empirical face-resistance factor of $N_0 = 2.5$ was adopted, consistent with recommended values for sandy soils (\citeblue{japan2000pipe, shou2010frictional}), as discussed in Section \blueeqref{deterministic}. For the $N_0$ estimates shown in Fig. \blueeqref{JP4_Comparison_Unlubricated}, all three methods eventually approach similar final values, although their trajectories across updates differ markedly. The MCMC-based Bayesian inference demonstrates relatively stable behavior, with 90\% credible intervals indicating the uncertainty associated with the estimates. After a slight decrease during the first update, the updated mean of $N_0$ gradually reaches approximately 2.3, with modest variations. In contrast, GA and PSO exhibit substantial fluctuations. Both methods undergo sharp declines during the early stages, with GA falling below 1.0 and PSO approaching zero within the first three updates. Both estimates subsequently recover, although with some fluctuations, reaching approximately 3.0 around 18 m before declining toward the MCMC-based mean.

For the residual interface friction coefficient $\mu$, the MCMC-based updated estimates show an initial decrease, followed by a gradual increase and eventual stabilization around 0.48. In comparison, GA and PSO both exhibit upward trends during the early stages, followed by divergent trajectories and pronounced fluctuations across subsequent updates. Around 15 m into the drive, the PSO estimate drops to near zero, followed by a sharp rebound above 0.5 in the next update. Across much of the drive, the GA and PSO estimates remain above the MCMC-based mean and the initial value, suggesting a tendency to overestimate $\mu$. From approximately 17 m onward, the PSO estimates become relatively stable, whereas GA continues to fluctuate noticeably. Overall, both methods display greater variability than the MCMC-based estimates, suggesting greater sensitivity to local data variations, while neither provides uncertainty quantification.
\begin{figure}[htbp]
    \centering
    \includegraphics[width=0.98\linewidth]{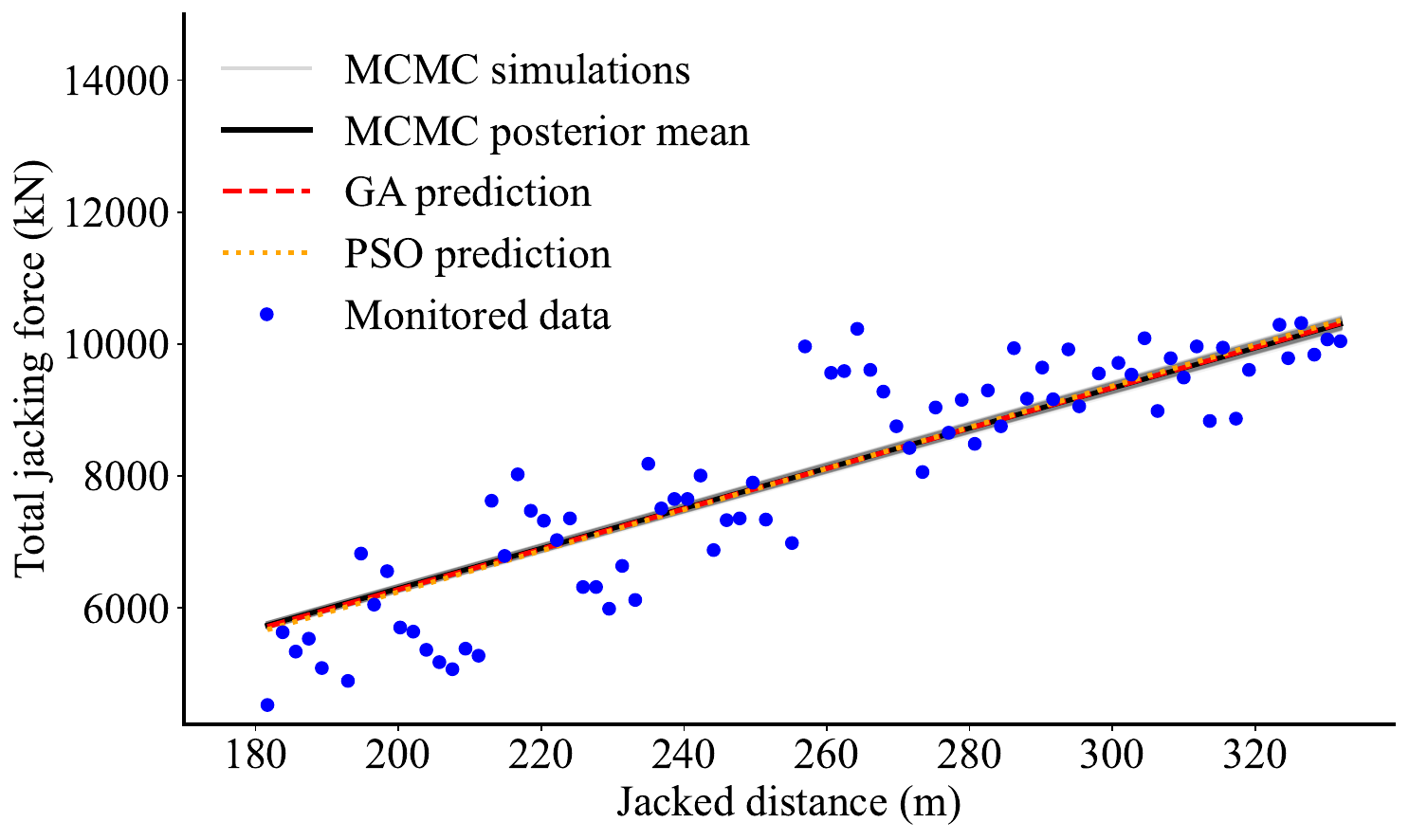} 
    \caption{\normalsize \textrm{Comparison of the reconstructed total jacking forces for the unlubricated segment of the Morris Avenue drive using MCMC-based Bayesian inference, GA, and PSO.}}
    \label{Morris_sim_comparison}
\end{figure}
\begin{figure*}[h]
    \centering
    \includegraphics[width=0.98\linewidth]{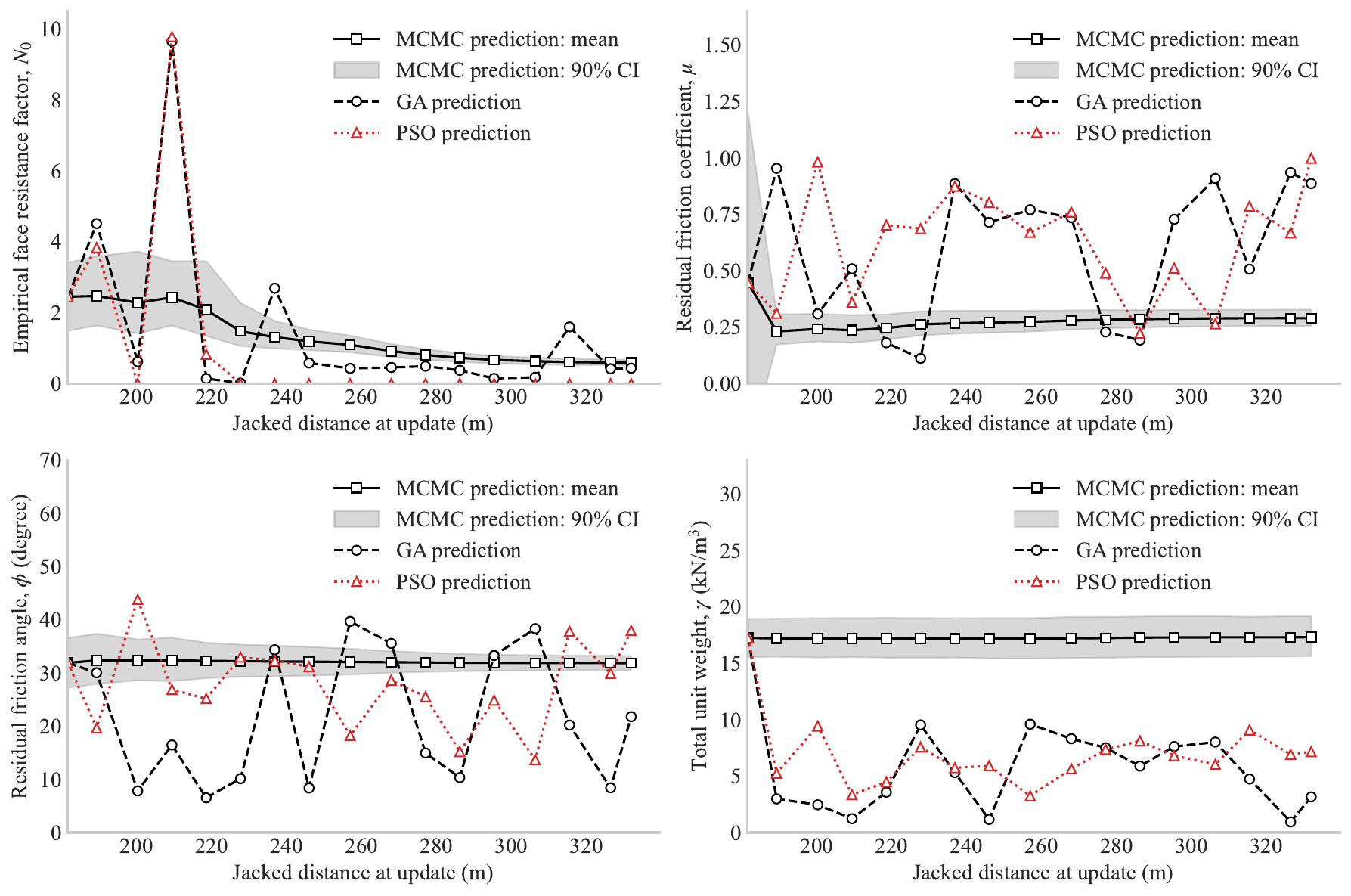} 
    \caption{\normalsize \textrm{Sequential estimates of the soil properties for the unlubricated segment of the Morris Avenue drive using MCMC-based Bayesian inference, GA, and PSO.}}
    \label{Morris_Comparison_Unlubricated}
\end{figure*}
For the residual friction angle $\phi$ and total unit weight $\gamma$, the MCMC-based estimates remain close to their initial values, with 90\% credible intervals indicating the uncertainty associated with the estimates. In contrast, both GA and PSO again show a sharp drop in the early updates, followed by lower estimates and pronounced fluctuations across subsequent updates. 

\subsection{Morris Avenue Tunnel}
The Morris Avenue Tunnel project, located in Renton, Washington, involved the installation of approximately 340 m of reinforced concrete pipe using microtunneling. Based on the site-investigation information shown in Fig. \blueeqref{Borehole_Combined}, the ground near the reception shaft consisted predominantly of silty sand. The pipe had an outer diameter of 2.22 m and an inner diameter of 1.83 m. Excavation was performed using a LovatMTS microtunneling machine with an outer diameter of 2.25 m, resulting in a diameter overcut of approximately 0.03 m. Automated bentonite lubrication was applied from the beginning of the drive up to approximately 179.80 m. The current analysis focuses on the segment beginning at approximately 180 m, after lubrication ceased.

As shown in Fig.~\blueeqref{Morris_sim_comparison}, all three methods reproduce the observed total jacking force trend with comparable agreement, yielding $R^2$ values ranging from 0.79 to 0.80. The MCMC-based approach additionally quantifies uncertainty in the reconstructed forces, illustrated by the spread around the mean estimate of the force simulations generated from the updated parameter distributions. Despite the similarity in reconstructed forces, the resulting updates to the model parameters $\theta$ differ substantially across methods, demonstrating that similar force reconstructions can correspond to different parameter combinations.

\begin{table*}[h]
    \centering
    {\fontfamily{ptm}\selectfont
    \caption{\normalsize \textrm{Initial parameter values and MCMC-based estimates for the microtunneling case studies}}
    \label{mcmc_inferences}
    \renewcommand{\arraystretch}{1.05}
    \begin{tabular}{lccccccc}
        \toprule
        \multirow{2}{*}{\textbf{Parameter}}
            & \multicolumn{3}{c}{\textbf{Alvarado JP3--RP4}}
            & \multicolumn{2}{c}{\textbf{Alvarado JP4--RP4}}
            & \multicolumn{2}{c}{\textbf{Morris Avenue}} \\
        \cmidrule(lr){2-4} \cmidrule(lr){5-6} \cmidrule(lr){7-8}
            & Prior & MCMC & MCMC (lub)
            & Prior & MCMC
            & Prior & MCMC \\
        \midrule
        Residual Soil friction angle $\phi$ (\si{\degree})
            & 31    & 30.7  & 32.63
            & 30    & 29.7
            & 32    & 31.8 \\
        Total Soil unit weight $\gamma$ (\si{kN/m^3})
            & 21    & 20.6  & 20.15
            & 20    & 20.4
            & 17    & 17.2 \\
        Empirical face resistance factor $N_0$
            & 2.5   & 2.2   & 42.5
            & 2.5   & 2.3
            & 2.5   & 1.0 \\
        Interface friction coefficient $\mu$
            & 0.49  & 0.50  & 0.05
            & 0.45  & 0.47
            & 0.59  & 0.29 \\
        Model and measurement error $\epsilon$ (kN)
            & 100   & 60    & 182
            & 100   & 65
            & 100   & 503 \\
        \bottomrule
    \end{tabular}

    \vspace{0.5em}
    \begin{minipage}{0.9\textwidth}
    \footnotesize
    \textbf{MCMC} only indicates the posterior estimates inferred from unlubricated segments.
    \end{minipage}
    }
\end{table*}

In this study, an initial empirical face resistance factor of $N_0 = 2.5$ was adopted, consistent with recommended values for sandy soil. As shown in Fig. \blueeqref{Morris_Comparison_Unlubricated}, the MCMC-based mean of $N_0$ remains close to the initial value over the first 30 meters of the analyzed segment, with relatively wide 90\% credible intervals indicating higher uncertainty in the early stage. As additional observations are incorporated, the estimates gradually decrease and stabilize around 1.0 during the later updates. In addition, the narrowing credible intervals indicate reduced uncertainty in the inference. The reduction in $N_0$ from its initial value of 2.5 is consistent with the silty-sand conditions identified for this case, whereas the initial value represents clean sandy soils. Earlier bentonite lubrication may also have influenced the inferred values. As previously noted, the current analysis focuses on the unlubricated segment of the drive, beginning at approximately 180 m. Prior to this point, bentonite lubrication was applied to reduce interface friction, which likely led to lower observed jacking forces and, consequently, affected the inferred values of $N_0$. In contrast, the estimates obtained using GA and PSO show pronounced fluctuations during the early updates, underscoring their relative instability. Although both initially follow broadly similar trends, their estimates diverge beyond approximately 230 m. From this point onward, the PSO estimates remain at the lower bound near zero, failing to adapt to new information during subsequent updates. The GA estimates remain relatively low until approximately 310 m, followed by a sharp increase and another decline toward the end of the drive.

For the residual interface friction coefficient $\mu$, the MCMC-based estimates decrease after the initial update and then stabilize around 0.3, with relatively narrow 90\% credible intervals during the subsequent updates. Although this final value deviates from the initial value, it remains consistent with field expectations. The reduction from the initial value may be associated with the earlier application of bentonite along the drive, as its friction-reducing effects at the pipe–soil interface could continue to influence the observed jacking forces, thereby lowering the inferred $\mu$ even after lubrication ceased. In contrast, the estimates produced by GA and PSO exhibit considerable fluctuations throughout the drive and end above the initial value, not converging to physically reasonable values given the expected influence of lubrication. This suggests that their stepwise optimization, without carrying forward updated parameter distributions, may increase sensitivity to local data variations and limit their ability to capture consistent parameter trends. Moreover, their lack of uncertainty quantification limits the interpretation of parameter behavior across the drive compared with the probabilistic method.

The MCMC-based estimates of the residual friction angle $\phi$ and total unit weight $\gamma$ remain close to their initial values, with 90\% credible intervals indicating the uncertainty associated with the estimates. In contrast, the GA and PSO estimates fluctuate considerably along the drive and remain lower on average than the initial values. These trends are consistent with the patterns observed in Figs. \blueeqref{JP3_Comparison_Unlubricated} and \blueeqref{JP4_Comparison_Unlubricated}, demonstrating stable estimation of these weakly identifiable parameters through the constraints imposed by the initial distributions and their relatively small COVs.
\section{Lubricated case study}
The results presented in Section \blueeqref{case} demonstrated that, across three unlubricated microtunneling drives, the proposed MCMC-based Bayesian inference framework provides parameter estimates that are more stable and consistent with field expectations than those obtained using GA and PSO, while explicitly quantifying uncertainty. To further evaluate its applicability under lubricated conditions, we apply the framework to a lubricated segment of the JP3--RP4 drive. This segment, between 118 m and 216 m, was subject to bentonite lubrication and lies within a homogeneous layer of medium-dense silty sand below the groundwater table.

\begin{figure*}[htbp]
    \centering
    \includegraphics[width=0.98\linewidth]{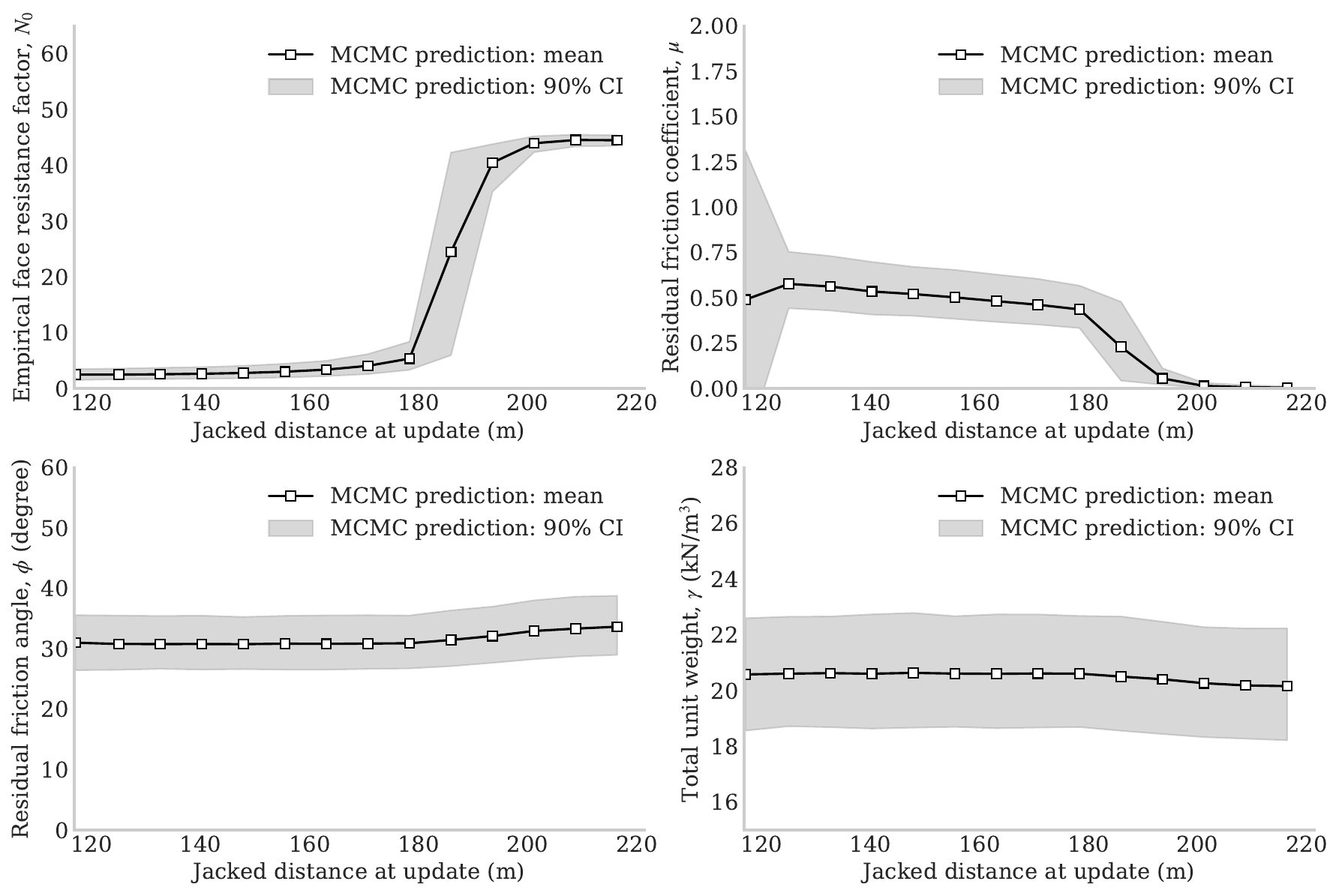} 
    \caption{\normalsize \textrm{Sequential estimates of the soil properties for the lubricated segment of the Alvarado Trunk Sewer JP3–RP4 drive using MCMC-based Bayesian inference.}}
    \label{JP3_Lubricated_Evolution}
\end{figure*}

Fig. \blueeqref{JP3_Lubricated_Evolution} illustrates the evolution of the inferred parameters along the lubricated segment. Both the empirical face-resistance factor $N_0$ and the residual interface friction coefficient $\mu$ exhibit substantial deviations from their initial values, with $\mu$ decreasing and stabilizing near 0.05 while $N_0$ increases markedly to approximately 42. In the lubricated case study, the initial parameter distributions used in the inference were derived from site-investigation information characterizing the native soil properties, whereas the observations reflect the altered pipe--soil interaction introduced by bentonite slurry. As these observations are incorporated, the inferred parameters depart from their initial values to reflect the effective soil behavior under lubricated conditions. In particular, the decrease in $\mu$ is consistent with the reduction in interface friction expected under lubrication, demonstrating that the framework incorporates information from observations and adapts its estimates beyond the initial assumptions. Table \blueeqref{mcmc_inferences} summarizes the initial parameter values and the MCMC-based estimates for the unlubricated and lubricated segments.

Lubrication applied along the pipe reduces resistance at the pipe--soil interface without acting directly on the cutting face, and the empirical face-resistance factor is expected to remain relatively constant under homogeneous soil conditions. Therefore, pipe lubrication alone would not explain the substantial increase in $N_0$ and we propose that this change may be associated with the following reasons. First, the conventional empirical values of $N_0$ (1.0 for clayey soils, 2.5 for sandy soils, and 3.0 for gravelly soils)  may not fully represent the effective face-resistance factor inferred under lubricated conditions. Further calibration using data from additional lubricated case studies may therefore be needed to assess and refine these reference values. Second, parameter compensation may occur within the proposed inference framework. The current single-layer formulation uses one likelihood based on total jacking force and does not independently model the latent lubrication effects. Consequently, it may not be able to decompose the observed total force into face resistance, pipe--soil friction, and lubrication contributions. An increase in modeled face resistance can offset a decrease in modeled frictional resistance while maintaining agreement with the observed total force. The simultaneous increase in $N_0$ and decrease in $\mu$ may reflect this compensation, highlighting the practical ill-posedness of decomposing the force components using the current formulation. Although the framework adapts its parameters to the observed response, the physical interpretation of the individual estimates remains uncertain under lubricated conditions. This limitation motivates extensions that explicitly represent lubrication effects and introduce additional information or physical constraints to improve the separation of the force components.

\section{Discussion}
The comparative analyses across three unlubricated drives from two projects demonstrate the applicability of the proposed Bayesian inference framework using the No-U-Turn Sampler (NUTS). Compared with Genetic Algorithm (GA) and Particle Swarm Optimization (PSO), the framework produces parameter estimates that are more stable and consistent with field expectations while explicitly quantifying uncertainty. The cases encompass different tunnel alignments and relatively homogeneous soil layers, supporting the applicability of the framework across different site conditions.

Comparisons of force reconstructions across the three methods show substantial differences in the inferred parameters, demonstrating that force agreement alone is insufficient to assess geotechnical-property characterization. The Bayesian framework combines site-investigation-based initial distributions with monitored forces to update parameter distributions and quantify uncertainty beyond point estimates. For the empirical face-resistance factor $N_0$ and residual interface friction coefficient $\mu$, the inferences generally become more stable as additional observations are incorporated, while the narrowing 90\% credible intervals indicate reduced uncertainty. For the weakly identifiable parameters $\phi$ and $\gamma$, the relatively small initial COVs constrain the adjustments and support stable inference. By combining the initial distributions with information from observations, the model adapts across different sites while maintaining physical plausibility in its inferences. The sequential inference structure further allows the model to track the spatial evolution of the inferred parameters along the tunnel alignment, offering continuous subsurface characterization that complements discrete site investigations.

In contrast, although some final estimates from GA and PSO approach the MCMC-based means or the initial parameter values, their update trajectories exhibit pronounced fluctuations, particularly during the early stages of the drives. These fluctuations may partly arise from their stepwise optimization without the constraints provided by initial parameter distributions, making their estimates more sensitive to local data variations. As a result, changes in the observations or optimization outcomes can favor different parameter combinations that produce similar force reconstructions. In addition, predefined search bounds restrict exploration of the parameter space and may cause estimates to remain near the boundaries, limiting their ability to adapt to new observations. Together, these limitations are particularly consequential in geotechnical applications, where tunneling conditions can shift spatially even within relatively homogeneous soil domains.

The lubricated case further demonstrates that the Bayesian framework responds to observations rather than remaining fixed at its initial assumptions. The decrease in $\mu$ is consistent with reduced interface friction, indicating adaptation to the effective behavior observed under lubrication. However, the substantial changes in $N_0$ suggest that this adaptation may also involve compensation among parameters and force components. This behavior highlights a limitation of prior regularization when the forward model omits a relevant physical mechanism. The current formulation, based on a single likelihood for total jacking force without explicitly modeling latent lubrication effects, cannot by itself distinguish lubrication-induced force reductions from changes in the underlying geotechnical properties, even though it uses the initial distributions to constrain parameter inference. Parameter compensation may consequently preserve agreement with measured forces while limiting decomposition into face resistance, native pipe–soil friction, and lubrication contributions and undermining the physical interpretation of the individual estimates. 


Future work will therefore explore extensions that explicitly model lubrication effects to better decompose the force components and reduce parameter compensation. Introducing latent lubrication variables, together with additional observations and physical constraints, may help distinguish lubrication-induced force reductions from changes in the underlying soil parameters. Probabilistic machine learning approaches, such as Bayesian Neural Networks (BNNs) and Gaussian Mixture Models, may also help capture the complex, nonlinear behavior introduced by bentonite while retaining uncertainty quantification.
\enlargethispage{\baselineskip}
\section{Conclusions}
This study proposed a Bayesian sequential inference framework for geotechnical-property characterization in microtunneling, integrating data preprocessing and sequential parameter updating with explicit uncertainty quantification and using initial distributions derived from site-investigation information to regularize the ill-posed inversion of monitored jacking forces. Controlled synthetic experiments demonstrated parameter recovery under varying noise levels and distinct responses to measurement noise and forward-model errors. Across three unlubricated drives, all three methods achieved comparable force-reconstruction performance, whereas the Bayesian framework produced more stable and physically interpretable parameter estimates consistent with field expectations by constraining the inference toward physically plausible parameter combinations. The lubricated case further demonstrated adaptation to observations beyond the initial assumptions, but also highlighted possible compensation among parameters and force components. The current single-layer likelihood formulation does not independently model latent lubrication effects, limiting force decomposition and parameter identification and reintroducing non-unique parameter combinations when the forward model omits the physics generating the observations. Future work will therefore explore explicit lubrication modeling through latent variables or hierarchical formulations, supported by additional observations and physical constraints, to improve force decomposition and reduce parameter compensation while retaining uncertainty quantification.

\section*{Data availability}
\noindent
The field monitoring records and site-investigation data used in this study were obtained from industry project partners and are subject to confidentiality
restrictions, so they cannot be made publicly available. The synthetic datasets and inference code are available from the corresponding author on reasonable request.
\section*{CRediT authorship contribution statement}
\noindent
\textbf{Honglai Peng:} Conceptualization, Methodology, Software, Formal analysis,
Investigation, Data curation, Visualization, Writing -- original draft.
\textbf{Rodrigo Borela:} Conceptualization, Validation, Supervision, Writing -- review \& editing.
\textbf{Kimberlie Staheli:} Resources, Data curation, Writing -- review \& editing.
\textbf{J. David Frost:} Conceptualization, Supervision, Funding acquisition,
Project administration, Writing -- review \& editing.

\section*{Acknowledgements}
\noindent
The findings presented in this manuscript are based on work supported in part by the US National Science Foundation through PTE Federal Award No. EEC-1449501. Any opinions, findings, and conclusions or recommendations expressed in this material are those of the authors and do not necessarily reflect the views of the National Science Foundation. Additional funding was provided by the Elizabeth and Bill Higginbotham Professorship in the School of Civil and Environmental Engineering at Georgia Tech.

\section*{Declaration of generative AI and AI-assisted technologies in the manuscript preparation process}
\noindent
During the preparation of this work, the authors used ChatGPT to assist with rephrasing and to improve the clarity and readability of the manuscript. After using this tool, the authors reviewed and edited the content as needed and take full responsibility for the content of the published article.





\renewcommand{\bibfont}{\normalsize} 
\renewcommand\refname{References}
\printbibliography

@book{french2010microtunneling,
  title={Microtunneling and Horizontal Drilling: Recommendations},
  author={{French Society for Trenchless Technology (FSTT)}},
  year={2010},
  publisher={John Wiley \& Sons}
}

@inproceedings{phillips2019instrumentation,
  title={Instrumentation and monitoring of a concrete jacking pipe},
  author={Phillips, BM and Royston, R and Sheil, BB and Byrne, BW},
  booktitle={International Conference on Smart Infrastructure and Construction 2019 (ICSIC) Driving data-informed decision-making},
  pages={457--462},
  year={2019},
  organization={ICE Publishing}
}

@article{mariethoz2010bayesian,
  title={Bayesian inverse problem and optimization with iterative spatial resampling},
  author={Mariethoz, Gr{\'e}goire and Renard, Philippe and Caers, Jef},
  journal={Water Resources Research},
  volume={46},
  number={11},
  year={2010},
  publisher={Wiley Online Library}
}

@article{calvetti2018inverse,
  title={Inverse problems: From regularization to Bayesian inference},
  author={Calvetti, Daniela and Somersalo, Erkki},
  journal={Wiley Interdisciplinary Reviews: Computational Statistics},
  volume={10},
  number={3},
  pages={e1427},
  year={2018},
  publisher={Wiley Online Library}
}

@article{ching2019constructing,
  title={Constructing site-specific multivariate probability distribution model using Bayesian machine learning},
  author={Ching, Jianye and Phoon, Kok-Kwang},
  journal={Journal of Engineering Mechanics},
  volume={145},
  number={1},
  pages={04018126},
  year={2019},
  publisher={American Society of Civil Engineers}
}

@article{cheng2023face,
  title={Face stability of shallowly buried large-section EPB box jacking crossing the Beijing-Hangzhou Grand Canal},
  author={Cheng, Cheng and Yang, Hao and Jia, Pengjiao and Ni, Pengpeng and Shi, Peixin and Ma, Peng and Xiang, Qiao},
  journal={Tunnelling and Underground Space Technology},
  volume={138},
  pages={105200},
  year={2023},
  publisher={Elsevier}
}

@article{shou2010frictional,
  title={On the frictional property of lubricants and its impact on jacking force and soil--pipe interaction of pipe-jacking},
  author={Shou, K and Yen, J and Liu, M},
  journal={Tunnelling and Underground Space Technology},
  volume={25},
  number={4},
  pages={469--477},
  year={2010},
  publisher={Elsevier}
}

@article{japan2000pipe,
  title={Pipe-jacking application},
  author={{JMTA}},
  journal={JMTA, Tokyo},
  year={2000}
}

@book{staheli2006jacking,
  title={Jacking force prediction: An interface friction approach based on pipe surface roughness},
  author={Staheli, Kimberlie},
  year={2006},
  publisher={Georgia Institute of Technology}
}

@article{doi:10.1061/AJRUA6.0000992,
author = {Kok-Kwang Phoon },
title = {Probabilistic Site Characterization},
journal = {ASCE-ASME Journal of Risk and Uncertainty in Engineering Systems, Part A: Civil Engineering},
volume = {4},
number = {4},
pages = {02018002},
year = {2018},

}

@inproceedings{phoon2019managing,
  title={Managing risk in geotechnical engineering--from data to digitalization},
  author={Phoon, Kok-Kwang and Ching, Jianye and Wang, Yu},
  booktitle={Proc., 7th Int. Symp. on Geotechnical Safety and Risk (ISGSR 2019)},
  pages={13--34},
  year={2019}
}

@misc{yuen2021bayesian,
  title={Bayesian learning methods for geotechnical data},
  author={Yuen, Ka-Veng and Ching, Jianye and Phoon, Kok-Kwang},
  year={2021},
  publisher={American Society of Civil Engineers}
}

@article{sheil2020machine,
  title={Machine learning to inform tunnelling operations: Recent advances and future trends},
  author={Sheil, Brian B and Suryasentana, Stephen K and Mooney, Michael A and Zhu, Hehua},
  journal={Proceedings of the Institution of Civil Engineers-Smart Infrastructure and Construction},
  volume={173},
  number={4},
  pages={74--95},
  year={2020},
  publisher={Thomas Telford Ltd}
}

@misc{sheil2020prediction,
  title={Prediction of microtunnelling jacking forces using a probabilistic observational approach. Tunn Undergr Sp Technol 109: 103749},
  author={Sheil, B},
  year={2020}
}

@article{sheil2022prediction,
  title={Prediction of pipe-jacking forces using a Bayesian updating approach},
  author={Sheil, Brian and Suryasentana, Stephen K and Templeman, Jack O and Phillips, Bryn M and Cheng, Wen-Chieh and Zhang, Limin},
  journal={Journal of Geotechnical and Geoenvironmental Engineering},
  volume={148},
  number={1},
  pages={04021173},
  year={2022},
  publisher={American Society of Civil Engineers}
}

@article{yu2024bayesian,
  title={Bayesian updating of model parameters for jacking force prediction using field observations},
  author={Shimada, Hideki, and Yu, Bosong, and Sasaoka, Takashi and Hamanaka, Akihiro and Matsumoto, Fumihiko and Morita, Tomo},
  journal={Tunnelling and Underground Space Technology},
  volume={149},
  pages={105810},
  year={2024},
  publisher={Elsevier}
}

@article{yu2024prediction,
  title={Prediction of Box-Jacking Force Using a Probabilistic Observational Approach},
  author={YU, Bosong and SHIMADA, Hideki and SASAOKA, Takashi and HAMANAKA, Akihiro and MATSUMOTO, Fumihiko and MORITA, Tomo},
  journal={International Journal of the JSRM},
  volume={20},
  number={2},
  pages={1--12},
  year={2024},
  publisher={Japanese Society for Rock Mechanics}
}

@book{jaynes2003probability,
  title={Probability theory: The logic of science},
  author={Jaynes, Edwin T},
  year={2003},
  publisher={Cambridge university press}
}

@article{beck1998updating,
  title={Updating models and their uncertainties. I: Bayesian statistical framework},
  author={Beck, James L and Katafygiotis, Lambros S},
  journal={Journal of Engineering Mechanics},
  volume={124},
  number={4},
  pages={455--461},
  year={1998},
  publisher={American Society of Civil Engineers}
}

@misc{box2011bayesian,
  title={Bayesian inference in statistical analysis (Vol. 40)},
  author={Box, GE and Tiao, George C},
  year={2011},
  publisher={John Wiley \& Sons Hoboken}
}

@book{gelman1995bayesian,
  title={Bayesian data analysis},
  author={Gelman, Andrew and Carlin, John B and Stern, Hal S and Rubin, Donald B},
  year={1995},
  publisher={Chapman and Hall/CRC}
}

@book{murphy2012machine,
  title={Machine learning: a probabilistic perspective},
  author={Murphy, Kevin P},
  year={2012},
  publisher={MIT press}
}

@article{van2018simple,
  title={A simple introduction to Markov Chain Monte--Carlo sampling},
  author={Van Ravenzwaaij, Don and Cassey, Pete and Brown, Scott D},
  journal={Psychonomic bulletin \& review},
  volume={25},
  number={1},
  pages={143--154},
  year={2018},
  publisher={Springer}
}

@book{gamerman2006markov,
  title={Markov chain Monte Carlo: stochastic simulation for Bayesian inference},
  author={Gamerman, Dani and Lopes, Hedibert F},
  year={2006},
  publisher={Chapman and Hall/CRC}
}

@book{gilks1995markov,
  title={Markov chain Monte Carlo in practice},
  author={Gilks, Walter R and Richardson, Sylvia and Spiegelhalter, David},
  year={1995},
  publisher={CRC press}
}

@article{hoffman2014no,
  title={The No-U-Turn sampler: adaptively setting path lengths in Hamiltonian Monte Carlo.},
  author={Hoffman, Matthew D and Gelman, Andrew and others},
  journal={J. Mach. Learn. Res.},
  volume={15},
  number={1},
  pages={1593--1623},
  year={2014}
}

@article{cleveland6seasonal,
  title={A seasonal-trend decomposition procedure based on Loess (with discussion)},
  author={Cleveland, STL},
  year={1990},
  journal={J. Off. Stat},
  volume={6},
  number={3}
}

@book{hyndman2018forecasting,
  title={Forecasting: principles and practice},
  author={Hyndman, Rob J and Athanasopoulos, George},
  year={2018},
  publisher={OTexts}
}

@article{gnanadesikan1968probability,
  title={Probability plotting methods for the analysis of data},
  author={Gnanadesikan, Ramanathan and Wilk, Martin B},
  journal={Biometrika},
  volume={55},
  number={1},
  pages={1--17},
  year={1968}
}

@article{li2016assessment,
  title={Assessment of slope stability in the monitoring parameter space},
  author={Li, XY and Zhang, LM and Jiang, SH and Li, DQ and Zhou, CB},
  journal={Journal of Geotechnical and Geoenvironmental Engineering},
  volume={142},
  number={7},
  pages={04016029},
  year={2016},
  publisher={American Society of Civil Engineers}
}

@article{zheng2018embankment,
  title={Embankment prediction using testing data and monitored behaviour: A Bayesian updating approach},
  author={Zheng, Dong and Huang, Jinsong and Li, Dian-Qing and Kelly, Richard and Sloan, Scott William},
  journal={Computers and Geotechnics},
  volume={93},
  pages={150--162},
  year={2018},
  publisher={Elsevier}
}

@inproceedings{kennedy1995particle,
  title={Particle swarm optimization},
  author={Kennedy, James and Eberhart, Russell},
  booktitle={Proceedings of ICNN'95-international conference on neural networks},
  volume={4},
  pages={1942--1948},
  year={1995},
  organization={ieee}
}

@article{hajihassani2018applications,
  title={Applications of particle swarm optimization in geotechnical engineering: a comprehensive review},
  author={Hajihassani, M and Jahed Armaghani, D and Kalatehjari, R},
  journal={Geotechnical and Geological Engineering},
  volume={36},
  pages={705--722},
  year={2018},
  publisher={Springer}
}

@article{zhou2023prediction,
  title={Prediction of jacking force using PSO-BPNN and PSO-SVR algorithm in curved pipe roof},
  author={Zhou, Hao and Huang, Sheng and Zhang, Peng and Ma, Baosong and Ma, Peng and Feng, Xin},
  journal={Tunnelling and Underground Space Technology},
  volume={138},
  pages={105159},
  year={2023},
  publisher={Elsevier}
}

@article{bo2024prediction,
  title={Prediction of tunnel deformation using PSO variant integrated with XGBoost and its TBM jamming application},
  author={Bo, Yin and Guo, Xiaogang and Liu, Quansheng and Pan, Yucong and Zhang, Lei and Lu, You},
  journal={Tunnelling and Underground Space Technology},
  volume={150},
  pages={105842},
  year={2024},
  publisher={Elsevier}
}

@article{jiang2011feedback,
  title={Feedback analysis of tunnel construction using a hybrid arithmetic based on support vector machine and particle swarm optimisation},
  author={Jiang, AN and Wang, SY and Tang, SL},
  journal={Automation in Construction},
  volume={20},
  number={4},
  pages={482--489},
  year={2011},
  publisher={Elsevier}
}

@article{holland1992genetic,
  title={Genetic algorithms},
  author={Holland, John H},
  journal={Scientific american},
  volume={267},
  number={1},
  pages={66--73},
  year={1992},
  publisher={JSTOR}
}

@article{andrab2017review,
  title={A review: evolutionary computations (GA and PSO) in geotechnical engineering},
  author={Andrab, Syed Gous and Hekmat, Ali and Yusop, Zulkifli Bin},
  journal={Computational Water, Energy, and Environmental Engineering},
  volume={6},
  number={2},
  pages={154--179},
  year={2017},
  publisher={Scientific Research Publishing}
}

@article{levasseur2008soil,
  title={Soil parameter identification using a genetic algorithm},
  author={Levasseur, S{\'e}verine and Mal{\'e}cot, Yann and Boulon, Marc and Flavigny, Etienne},
  journal={International journal for numerical and analytical methods in geomechanics},
  volume={32},
  number={2},
  pages={189--213},
  year={2008},
  publisher={Wiley Online Library}
}

@article{vardakos2012parameter,
  title={Parameter identification in numerical modeling of tunneling using the Differential Evolution Genetic Algorithm (DEGA)},
  author={Vardakos, Sotirios and Gutierrez, Marte and Xia, Caichu},
  journal={Tunnelling and underground space technology},
  volume={28},
  pages={109--123},
  year={2012},
  publisher={Elsevier}
}

@article{feng2022enhanced,
  title={Enhanced prediction intervals of tunnel-induced settlement using the genetic algorithm and neural network},
  author={Feng, Liuyang and Zhang, Limao},
  journal={Reliability Engineering \& System Safety},
  volume={223},
  pages={108439},
  year={2022},
  publisher={Elsevier}
}

@article{harr1987reliability,
  title={Reliability-based design in civil engineering},
  author={Harr, Milton Edward},
  journal={(No Title)},
  year={1987}
}

@inproceedings{kulhawy1992evaluation,
  author    = {Fred H. Kulhawy},
  title     = {On the Evaluation of Static Soil Properties},
  booktitle = {ASCE Geotechnical Special Publication},
  volume    = {31},
  pages     = {95--115},
  year      = {1992},
  publisher = {ASCE}
}

@article{o2020interpretation,
  title={Interpretation of pipe-jacking and lubrication records for drives in silty soil},
  author={O'Dwyer, Kevin G and McCabe, Bryan A and Sheil, Brian B},
  journal={Underground Space},
  volume={5},
  number={3},
  pages={199--209},
  year={2020},
  publisher={Elsevier}
}

\end{document}